\documentclass[12pt]{article}

\AtBeginDocument{
  \addtocontents{toc}{\normalsize}
}
\pdfoutput=1

\usepackage{amsmath,amssymb,amsfonts,amsthm,amscd,mathrsfs}
\usepackage{xcolor}
\definecolor{darkblue}{rgb}{0.1,0.1,.7}
\usepackage[]{version}
\usepackage[]{graphicx}
\usepackage[]{latexsym}
\usepackage{geometry}
\usepackage[all,cmtip]{xy}

\usepackage[margin=10pt,font=small,labelfont=bf]{caption}
\usepackage{ifthen}
\usepackage{soul}
\usepackage{tikz}
\usepackage{array,setspace,mathrsfs,amsfonts,yfonts,dsfont,bbm,colonequals}
\usepackage{dsfont}
\usepackage{cite}
\usepackage{xspace}
\usepackage{empheq}

\usepackage{xcolor}
\usepackage[colorlinks, linkcolor=darkblue, citecolor=darkblue, urlcolor=darkblue, linktocpage]{hyperref} 

\usepackage{subcaption}
\usepackage[utf8]{inputenc}
\usepackage{setspace}
\usepackage{float, graphicx}

\numberwithin{equation}{section}

\newcommand{\mm}{a}
\newcommand{\te}{t}
\newcommand{\al}{(a\lambda)}
\newcommand{\hD}{{\hat \Delta}}
\newcommand{\hG}{{\widehat G}}
\newcommand{\ve}{\varepsilon}

\newcommand{\cO}{\mathcal O}

\newcommand{\reef}[1]{(\ref{#1})}
\newcommand{\be}{\begin{equation}}
\newcommand{\ee}{\end{equation}}
\newcommand{\bea}{\begin{eqnarray}}
\newcommand{\eea}{\end{eqnarray}}
\newcommand{\ba}{\begin{equation}\begin{aligned}}
\newcommand{\ea}{\end{aligned}\end{equation}}

\newcommand{\ud}{\mathrm d}

\newcommand{\Df}{{\Delta_\phi}}

\newcommand*\widefbox[1]{\fbox{\hspace{2em}#1\hspace{2em}}}

\def\d{\delta}

\def\t{\theta}
\def\D{\Delta}
\def\G{\Gamma}

\def\la{\langle}
\def\ra{\rangle}

\def\G{\Gamma}

\begin{document}

\vspace*{-.6in} \thispagestyle{empty}
\begin{flushright}
LPTENS/18/18
\end{flushright}
\vspace{1cm} {\Large
\begin{center}
{\bf The Functional Bootstrap for Boundary CFT}\\
\end{center}}
\vspace{1cm}
\begin{center}
{\bf Apratim Kaviraj${}^{D,N} $ , Miguel F.~Paulos${}^{D}$ }\\[1cm] 
{
\small
{\em  ${}^D$Laboratoire de Physique Th\'eorique \&  ${}^N$Institut de Physique Th\'{e}orique Philippe Meyer, \\ de l'\'Ecole Normale Sup\'erieure, PSL University,\\ CNRS, Sorbonne Universit\'es, UPMC Univ. Paris 06\\ 24 rue Lhomond, 75231 Paris Cedex 05, France
}
\normalsize
}
\\
\end{center}

\begin{center}
	{\texttt{apratim.kaviraj@lpt.ens.fr,} \ \  \texttt{miguel.paulos@lpt.ens.fr} 
	}
	\\
\end{center}

\vspace{4mm}

\begin{abstract}
We introduce a new approach to the study of the crossing equation for CFTs in the presence of a boundary. We argue that there is a basis for this equation related to the generalized free field solution. The dual basis is a set of linear functionals which act on the crossing equation to give a set of sum rules on the boundary CFT data: the functional bootstrap equations. We show these equations are essentially equivalent to a Polyakov-type approach to the bootstrap of BCFTs, and show how to fix the so-called contact term ambiguity in that context. Finally, the functional bootstrap equations diagonalize perturbation theory around generalized free fields, which we use to recover the Wilson-Fisher BCFT data in the $\epsilon$-expansion to order $\epsilon^2$.

\end{abstract}
\vspace{2in}


\newpage

{
\setlength{\parskip}{0.05in}
\tableofcontents
\renewcommand{\baselinestretch}{1.0}\normalsize
}


\setlength{\parskip}{0.1in}
\newpage

\section{Introduction}\label{sec:introduction}

The bootstrap philosophy \cite{Ferrara:1973yt,Polyakov:1974gs} was first applied to boundary conformal field theories (BCFTs) in \cite{Liendo2013} building on general results of \cite{McAvity:1995zd, CARDY1984514} and the methods introduced in \cite{Rattazzi:2008pe}. In that work, unitarity and crossing symmetry was used to derive numerical bounds for general BCFTs, as well as exact results for the Wilson-Fisher fixed points in the presence of a boundary. Since then, many other works have studied and generalized this to conformal defects in a number of situations \cite{Gliozzi2015,Billo:2016cpy,Lemos:2017vnx,Liendo:2018ukf,Hogervorst:2017kbj,Gliozzi:2016cmg,Gadde:2016fbj,Lauria:2017wav}. 

The usual bootstrap starting point are the two point functions, which in the presence of a defect are non-trivial functions of one or more cross-ratios. A notable feature is that unlike in the usual CFT bootstrap, the crossing equation for such correlators involves coefficients which lack a definite sign, even assuming unitarity. This makes ordinary numerical methods harder or impossible to justify. Approaches based on sparseness of extremal solutions \cite{Gliozzi2013, El-Showk2018} have so far obtained limited, but interesting, results \cite{Gliozzi2013,Gliozzi2016}, which should be extended and explored.

Very recently, the works \cite{Mazac2018,Mazac:2016qev,Mazac:2018ycv} have introduced a new analytic approach to studying 1d CFT correlators, or more generally CFTs restricted to a line.\footnote{See also \cite{Mazac:2018qmi} for related work.} In this approach one introduces a complete set of linear functionals associated to the generalized free field solutions to crossing. They are the extremal functionals \cite{ElShowk:2012hu,El-Showk2018} that correspond to these solutions, and are associated to optimal bounds on the CFT data. By acting on the crossing equation with these functionals one obtains a set of sum rules that this data must satisfy. These sum rules are a complete reformulation of the original crossing equation, and have many nice properties. Among them they provide upper and lower bounds on the OPE data, diagonalize perturbation theory around generalized free solutions, and also allow a rigorous derivation of the (Mellin-)Polyakov bootstrap \cite{Polyakov:1974gs,Sen:2015doa,Gopakumar2017,Gopakumar2017a,Ghosh:2018bgd}. In particular, in this 1d context it fixes the polynomial (or contact-term) ambiguity, that limits the applicability of the Polyakov bootstrap in higher dimensions\cite{Dey:2017fab,Gopakumar:2018xqi}.

The success of this approach suggests we should attempt to generalize it to other settings. In this work we will initiate the study of the functional approach to the boundary bootstrap. This is particularly nice as our results will be immediately relevant for higher-dimensional CFTs in the presence of a boundary, providing a useful stepping stone towards a functional bootstrap analysis of the higher-d CFT crossing equation.

Concretely, in this paper we will explain how to construct suitable bases of linear functionals that act on the crossing equation for a scalar CFT correlator $\langle \phi \phi\rangle$ on a half-space. These functionals are associated to the generalized free boson CFT in the presence of a boundary with Neumann or Dirichlet boundary conditions. Technically the construction shares sufficient the features of the 1d CFT case that will allow us to make progress in a relatively straightforward manner.

We act with our set of functionals on the crossing equation to obtain a set of functional bootstrap equations contraining the OPE and BOE data of the BCFT. As we explain, these equations can be equivalently be thought of as arising from a Polyakov-type approach to bootstrapping the BCFT crossing equation. In particular, the functional actions on conformal blocks compute the conformal block decompositions of Witten diagrams in an AdS$_{d+1}$ geometry in the presence of a radially extended $d$-dimensional brane. The functional bootstrap equations then guarantee that two-point functions can be written in a manifestly crossing symmetric manner in terms of Polyakov blocks for BCFTs.

By perturbing the functional bootstrap equations around the generalized free solution we can continuously deform it to more interesting interacting solutions. In particular, if we assume the states in the spectrum to be essentially deformations of those of the generalized free case, apart from eventually some fixed number of states which we control, the deformation is uniquely fixed. This is the idea of extremality: sparseness of a solution to crossing can be used to uniquely determine it. Here we will use this approach to rederive Wilson-Fisher BCFT data up to $O(\epsilon^2)$ \cite{Bissi2018}.

The outline of this paper is as follows. In the next section we briefly review the key points of the boundary bootstrap which are relevant to this work. Section \ref{sec:funcans} introduces the general ansatz for the functionals we will consider, and constructs a simpler, pedagogical basis of such functionals which can be used to derive a toy version of the functional bootstrap equations and the Polyakov bootstrap. The construction of the main functional basis takes place in section \ref{sec:functionals}, which is motivated by considering Witten diagrams in AdS space in the presence of a brane. This picture is confirmed in \ref{sec:witten} by explicitly computing the relevant Witten diagrams, where we show the functional actions appear as coefficients of bulk double trace operators and boundary derivative operators. In section \ref{sec:polyakov} we explain in detail how the functionals we construct are related to a Polyakov-like approach to the bootstrap of BCFT correlators. As an application of the functional bootstrap equations, we study them in perturbation theory around free theory in $4-\epsilon$ dimensions to recover the Wilson-Fisher BCFT OPE and BOE data to $O(\epsilon^2)$. We conclude with a discussion and outlook. The paper is complemented by technical appendices.
\vspace{0.1 cm}

\textbf{Note:} While this work was being completed we became aware of the work \cite{RMZ} which overlaps with ours.

\section{Review of the boundary bootstrap}
\label{sec:review}
In this section we will briefly review the bootstrap approach to BCFTs. We will be brief, referring the reader to the literature for further details, e.g. \cite{Liendo2013,McAvity:1995zd, CARDY1984514}.

We consider $d$-dimensional CFTs on the Euclidean half-space with coordinates:
\bea
(\vec x,x^\perp)=(x^1,\ldots x^{d-1},x^\perp)\,,\qquad \text{with}\quad x^\perp\geq 0.
\eea 
A boundary CFT arises when it is possible to choose boundary conditions at $x^\perp=0$ in such a way that a $SO(1,d)$ subgroup of the conformal group is unbroken, namely the group of conformal transformations which preserve the hyperplane $x^\perp=0$, which we will call the ``defect''. Under these circumstances, the theory contains both the original bulk operators $\cO$ as well as operators $\hat \cO$ which live on the defect. The latter can be thought of as arising by taking bulk operators sufficiently close to the defect. Correlators containing only defect operators satisfy all the axioms of an ordinary CFT with the exception of the existence of a stress-tensor.\footnote{Since the half-space is conformally invariant to AdS, this can be thought of as a special case of the construction in \cite{Paulos:2016fap}, where the bulk QFT happens to be a CFT.}

Since the original conformal group is partially broken, bulk scalar operators can now develop expectation values while preserving the $SO(1,d)$ subgroup:
\bea
\langle \cO(\vec x,x^\perp)\rangle=\frac{a_\cO}{(x^\perp)^{\Delta_\cO}}.
\eea
In turn this implies that the bulk two point function can become non-trivial. Indeed, with two points we can make up a cross-ratio which is invariant under the unbroken conformal transformations:\footnote{Note that this is related to the more usual $\xi=\frac{|\vec x_{12}|^2+(x_1^\perp-x_2^\perp)^2}{4x_1^\perp x_2^\perp}$ by $\xi=\frac{1-z}z$.}
\bea
z=\frac{4x_1^\perp x_2^\perp}{|\vec x_{12}|^2+(x_1^\perp+x_2^\perp)^2}, \qquad \vec x_{ij}\equiv \vec x_i-\vec x_j.
\eea
In terms of this cross-ratio we can write a scalar two-point function as
\bea
\langle \phi_1(\vec x_1,x_1^\perp)\phi_2(\vec x_2,x_2^\perp)\rangle=z^{\frac{\Delta_{\phi_1}+\Delta_{\phi_2}}2}\,\frac{\,\mathcal G(z)}{(x_1^\perp)^{\Delta_{\phi_1}}(x_2^\perp)^{\Delta_{\phi_2}}}.
\eea
We have pulled out a power of $z$ for convenience. In the bulk we have an OPE which takes the schematic form:
\bea
\phi_1 \times \phi_2=\sum_{\cO\in \phi_1\times \phi_2} \lambda_{\phi_1\phi_2 \cO} \left(\cO+\text{descendants}\right)
\eea
We can use this to derive an expansion for the correlator around $z=1$:
\bea
\mathcal G(z)= \sum_{\cO \in \phi_1\times \phi_2} a_{\cO} \lambda_{\phi_1 \phi_2 \cO}\, G_{\Delta_\cO}(z | \Delta_{\phi_1},\Delta_{\phi_2})\,,
\eea
where we have introduced the bulk conformal block,
\bea
G_{\Delta}(z|\Delta_{\phi_1},\Delta_{\phi_2})=\frac{(1-z)^{\frac{\Delta-\Delta_{\phi_1}-\Delta_{\phi_2}}2}}{z^{\frac{\Delta_{12}}2}}{}_2F_1\left(\frac{\Delta}2+\frac{\Delta_{12}}2,\frac \Delta 2-\varepsilon+\frac{\Delta_{12}}2,\Delta-\varepsilon,1-z\right),\label{eq:bulkblock}
\eea
with
\bea
\varepsilon\equiv \frac{d-2}2,\qquad  \Delta_{12}\equiv \Delta_{\phi_1}-\Delta_{\phi_2}
\eea
which captures contributions to the two-point function from a bulk primary and all its descendants.

It is possible to do a different decomposition of the same two point function \cite{Cardy:1991tv}. This arises from the fact that any bulk operator can be expanded in a basis of defect operators. This is a direct consequence of radially quantizing the theory around a point contained in the defect. This boundary operator expansion (BOE) takes the form
\bea
\phi(\vec x,x^\perp)=\sum_{\hat \cO} \mu_{\hat\cO}(x^\perp)^{\Delta_{\hat \cO}-\Df} C[(x^\perp)^2\partial^2] \hat \cO(\vec x)
\eea
where $C[(x^\perp)^2\partial^2]$ captures contributions of boundary descendants \cite{McAvity:1995zd}. Using the BOE we now have
\bea
\mathcal G(z)=\sum_{\hat \cO} \mu^2_{\hat \cO}  \hG_{\Delta_{\hat \cO}}(z|\Delta_{\phi_1},\Delta_{\phi_2})
\eea
where we've defined the boundary blocks:
\bea
\widehat G_{\hat \Delta}(z|\Delta_{\phi_1},\Delta_{\phi_2})=z^{\hat\Delta-\frac{\Delta_{\phi_1}+\Delta_{\phi_2}}2}{}_2F_1(\hD,\hD-\varepsilon,2\hD-2\varepsilon,z)\label{eq:bdyblock}.
\eea

From now on we will be interested in the case where $\Delta_{\phi_1}=\Delta_{\phi_2}=\Delta_\phi$. Equating the two different expansions for the correlator we obtain the boundary bootstrap crossing equation:
\bea
\boxed{
\sum_{\hat \Delta} \mu^2_{\hat \Delta} \widehat G_{\hat \Delta}(z|\Df)=\sum_{\Delta} (a \lambda)_\Delta G_{\Delta}(z|\Df).}\label{eq:crossingeq}
\eea
From now on we'll want to think of the above as a mathematical equation for the OPE and BOE data, whose solutions may or not correspond to a physical theory. Accordingly the sums now run over pure labels $\hat \Delta, \Delta$ and we have made several self-evident abreviations.

The BCFT crossing equation above is very similar to the $D=1$ CFT bootstrap crossing equation equating s- and t-channel expansions of a four-point function of identical operators. Indeed, for $\varepsilon=0$ and setting $(a\lambda)_{\Delta}=\mu^2_{\Delta/2}$ the two become equivalent. More generally however, not only are the two channels now independent, but crucially the bulk channel does not have positivity, since the $(a\lambda)_\Delta$ can have either sign. 

Although the original equation was defined only for $z\in(0,1)$ in the Euclidean region, we actually have convergence in the much wider region $\mathcal R\equiv \mathbb C\backslash (-\infty,0)\cup (1,\infty)$. This region is obtained by complexifying $z$ and avoiding the branch cuts of the bulk and boundary blocks.\footnote{It is possible to prove this in the same way as the analogous statement for the $D=1$ CFT, namely by introducing a radial coordinate $\rho(z)$ which maps the cut complex plane to the disk \cite{Hogervorst:2013sma}}

\subsubsection*{The generalized free solution}
An important example of a CFT in the presence of a boundary is the generalized free field (GFF) with Dirichlet or Neumann boundary conditions. This is a theory with an elementary field $\phi$ whose correlators are given by Wick contractions. The two point function in turn is given as
\bea
\langle \phi(\vec x_1,x_1^\perp)\phi(\vec x_2,x_2^\perp)\rangle&=&\frac{1}{(|\vec x_{12}|^2+(x_1^\perp-x_2^\perp)^2)^{\Df}}+\nu \frac{1}{(|\vec x_{12}|^2+(x_1^\perp+x_2^\perp)^2)^{\Df}}\nonumber \\
\Leftrightarrow \mathcal G(z)&=&\frac{1}{(1-z)^{\Df}}+\nu.
\eea
where $\nu=1,-1$ for Neumann, Dirichlet boundary conditions respectively. In the bulk channel the correlator contains the identity and double trace operators $\phi \Box^n \phi$ with dimension $\Delta_n=2\Df+2n$. In the boundary channel one finds operators with dimension $\hat \Delta_n^+=\Df+2n$ for Neumann boundary conditions, and $\hat \Delta_n^-=\Df+1+2n$ for Dirichlet, corresponding to (transverse) derivative operators $\partial_{\perp}^{2n} \phi(\vec x,0)$ or $\partial_{\perp}^{1+2n}\phi(\vec x,0)$ respectively. The OPE and BOE data in this case are \cite{Liendo2013}:
\begin{subequations}
\label{eq:gffdata}
\bea
(a\lambda)_{\Delta_n}&=&\nu (-1)^n\frac{(\Df)_n(2\Df-1-\ve+2n)_{-n}}{(\Df-\ve+n+1)_{-n}n!} \\
\mu_{\hat \Delta_n^+}^2&=&
 \frac{(\Df)_{2n} (\Df-\ve-\frac 12+2n)_{-n}}{2^{2n-1}(2n)!(\Df-\ve+n)_{-n}}, \qquad \mbox{Neumann}\\
 \mu_{\hat \Delta_n^-}^2&=&\frac{(\Df)_{2n+1} (\Df-\ve+\frac 12+2n)_{-n}}{2^{2n}(2n+1)!(\Df-\ve+n)_{-n}}, \qquad \mbox{Dirichlet}
\eea
\end{subequations}
The GFF theory can be thought of as a free scalar field with mass $m^2=\Df(\Df-d)$ living in AdS$_{d+1}$ \cite{RastelliMellin}. In the present context, we have placed a brane in AdS perpendicular to its boundary, partially breaking the AdS isometries. The induced metric on this brane is AdS$_{d}$, which makes manifest that a subgroup of conformal transformations is preserved. On this brane the field is chosen to have Neumann or Dirichlet boundary conditions.\footnote{This is not to be confused with the condition at the radial boundary of AdS$_{d+1}$.}
The resulting theory trivially reproduces the above correlators once AdS fields are pushed out to the boundary. It is natural however to think of deformations of the solution above, introduced by adding interactions in AdS$_{d+1}$. This will play an important role in our story later on.

\section{Functionals and a simple basis}\label{sec:funcans}

\subsection{Functional ansatz and constraints}
We propose that the optimal way to extract information from the crossing equation is to act on it with a suitable basis of linear functionals. The functionals should have the property that they commute with the infinite sums over states in the crossing equation -- the swapping condition \cite{Rychkov:2017tpc} -- and lead to finite results. Let us denote one such functional by $\omega$. Acting on the crossing equation we obtain a sum rule
\bea
\sum_{\hD} \mu^2_{\hD} \omega(\hD|\Df)=\sum_{\Delta} (a\lambda)_{\D} \omega(\D|\Df)\,,
\eea
which should be thought of as a constraint on the BCFT data. In the above we have defined the useful abbreviations:
\bea
\omega\left[\hG_{\hD}(z | \Df)\right]\equiv \omega(\hD|\Df),\qquad \omega\left[G_{\D}(z | \Df)\right]\equiv \omega(\D|\Df).
\eea
We ask for the indulgence of the reader for the slight abuse of notation, which will pay off in a lot less hats. We will also often drop the explicit $\Df$ dependence and use $\omega(\Delta), \omega(\hD)$.

The two basic properties we demand of a linear functional are:
\begin{itemize}
\item {\bf Finiteness}
\begin{equation}
\begin{aligned}
\omega(\hD)&<\infty \qquad \text{for}\quad \hD\geq 0\\
\omega(\D)&<\infty \qquad \text{for}\quad \D\geq 0\\
\omega[\mathcal G]&<\infty
\end{aligned}
\end{equation}
\item {\bf Swapping}
\begin{equation}
\begin{aligned}
\omega[\mathcal G]=\sum_{\hD} \mu^2_{\hD}\omega(\hD|\Df)=\sum_{\D} (a\lambda)_{\D} \omega(\D|\Df).
\end{aligned}
\end{equation}
\end{itemize}
These conditions are broad enough that they allow for many different sets of functionals. Clearly the original crossing constraints are recovered by considering the (trivial) set of functionals which evaluate a function at a specific point. A more interesting basis is the set of derivatives at $z=\frac 12$, which is the approach used in the numerical bootstrap~\cite{Rattazzi:2008pe}. In this paper our main focus is on an interesting set of functionals which in a sense are dual to the generalized free field solutions to crossing discussed at the end of the previous section. However, to illustrate the main ideas we will first consider a toy basis of functionals in the next subsection. 

For now, let us introduce our general ansatz for the functionals and examine in detail how it is constrained by the above requirements. A natural starting point is:
\bea
\omega[\mathcal F]:=\int_{-\infty}^{0} \ud z\, h_-(z) \mbox{Im}[\mathcal F(z)]+\int_{1}^{\infty} \ud z\,h_+(z) \mbox{Im}[\mathcal F(z)]\label{eq:ans1}
\eea
with $\mbox{Im}[\mathcal F(z)]:=\lim_{\epsilon\to 0^+}\frac{1}{2i}\left[ \mathcal F(z+i\epsilon)-\mathcal F^*(z-i\epsilon)\right]$.
It is very natural to consider such an ansatz, since we will be acting on functions which are holomorphic in the crossing region $\mathcal R$, and such functions are essentially determined by their values near the cuts $(-\infty,0)\cup (1,\infty)$. We can think of the above as integrating the kernels $h_+,h_-$ against test functions on a contour that wraps these cuts. Assuming for now sufficiently soft behaviour at $z=\infty$, we can deform the contour of integration to obtain instead:
\bea
\label{eq:funcactiondef}
\omega\left[\mathcal F\right]&=&\int_{\frac 12+i\mathbb R} \frac{\ud z}{2\pi i}\,[h_+(z)-h_-(z)] \mathcal F(z)\nonumber \\
&&+\int_{\frac 12}^1\frac{\ud z}\pi\mbox{Im}[h_+(z)] \mathcal F(z)-\int_{0}^{\frac 12} \frac{\ud z}\pi\mbox{Im}[h_-(z)] \mathcal F(z).\label{eq:funcans}
\eea
One should really think of this form of the functional action as the more fundamental one. In particular, when deforming the contour to go back to \reef{eq:ans1} we might not be allowed to wrap the contours around the branch points at $z=0,1$. 

From this expression we find that finiteness of the action on the conformal blocks \reef{eq:bulkblock}, \reef{eq:bdyblock} demands:
\begin{itemize}
\item{\bf Finiteness}:
\begin{subequations}
\label{eq:constraints1}
\bea
h_+(z)-h_-(z)&\overset{|z|\to \infty}{\sim}& O(|z|^{\Df-\ve-1-\eta}), \\
\mbox{Im}\left[ h_+(z)\right]&\overset{z\to 1^-}{\sim}& O[(1-z)^{\Df-1+\eta}],\\
\mbox{Im}\left[h_-(z)\right]&\overset{z\to 0^+}{\sim} &O[(-z)^{\Df-1+\eta}].
\eea
\end{subequations}
\end{itemize}
for some $\eta>0$.
Now consider swapping. Since the crossing sums are uniformly convergent in the crossing region, by deforming the contour of integration we see that we need only worry about the behaviour of the kernels in the regions near $\infty,0$ and $1$. However, it is clear that for the last two no additional constraints come from the swapping condition. For instance, it is true that the infinite sum over boundary blocks behaves worse near $z=1$ than any single term in the sum. However, we can use crossing to bound this behaviour by the bulk channel expansion, for which we are already assuming the functional action is finite. The same argument goes through mutatis mutandis for $z=0$. So, we must worry only about the behaviour at infinity. Here, there is indeed a potential problem, since the tail of the infinite sum over blocks falls off more slowly near $z=\infty$ than any individual block. Consider in particular the boundary channel. We have
\bea
\left|\sum_{\hD>\hD^*} \mu_\hD^2 \hG_{\hD}(z|\Df)\right|\leq \sum_{\hD>\hD^*} \mu_\hD^2 \frac{\hG_{\hD}\left(\left|\frac{z}{z-1}\right| |\Df\right)}{|1-z|^{\Df}}\underset{z\to \infty}{\leq} \frac{C}{|z-1|^{\Df}} \frac{1}{\left(1-\left|\frac{z}{z-1}\right|\right)^{\Df}}=O(1)
\eea
where in the last step we have used crossing and the fact that in this limit the leading contribution comes from the identity operator in the bulk channel. Using crossing, it is not hard to see that the tail of the sum over bulk blocks must be bounded in the same way. The contribution from this tail will be negligible as long as we make the contribution from the integration region near $z=\infty$ vanishingly small. This requires a stronger falloff condition on $h_+(z)-h_-(z)$. Given this condition, it is clear that also the action of the functional on the full correlator $\mathcal G(z)$ is also finite.

The conclusion is that overall, finiteness and swapping conditions constrain our ansatz in the following way:
\begin{itemize}
\item{\bf Finiteness+Swapping}
\begin{subequations}
\label{eq:constraints}
\bea
h_+(z)-h_-(z)&\overset{|z|\to \infty}{\sim}& O(|z|^{-1-\eta}), \\
\mbox{Im}\left[ h_+(z)\right]&\overset{z\to 1^-}{\sim}& O[(1-z)^{\Df-1+\eta}],\\
\mbox{Im}\left[h_-(z)\right]&\overset{z\to 0^+}{\sim} &O[(-z)^{\Df-1+\eta}].
\eea
\end{subequations}
\end{itemize}
The first condition above is the most non-trivial. Relaxing or strengthening it allows for functionals that act on objects with softer or harder behaviour near $z=\infty$. In particular, perturbative corrections to correlators may not share the same large $z$-behaviour as fully-fledged correlators.

\subsection{Warm-up: a simple basis}
\label{sec:easy}
\subsubsection{Special values of $\Df$ and orthonormal functionals}
We will now consider a simple basis of functionals which nevertheless has several nice properties. In this basis, the functionals will be dual to blocks in each channel with specific integer-spaced scaling dimensions that appear in the generalized free solutions. We will take a pedagogical route to explain general features which will be useful later, but the final result will be extremely simple.

The construction starts from the simple observation that the discontinuities of both bulk and boundary blocks are made up of two pieces. One of the pieces contains a sine function, which vanishes automatically for those values of $\Delta,\hat \Delta$ corresponding to the generalized free field solutions. The other piece is, for both cases, a Jacobi function of the first kind, which reduces to a polynomial when $\hD=1+\ve+n$ and $\Delta=2+2\ve+2n$ with $n$ integer. This suggests using the orthogonality properties of Jacobi polynomials to construct the following functionals:
\bea
\tau_{n}^{\mbox{\tiny pre}}:&& \qquad h_+(z)=0, \qquad h_-(z)\propto (1-z)^{\Df-2-\ve} P^{(0,\ve)}_{n}\left(\frac{1+z}{1-z}\right)\\
\widehat \tau_n^{\mbox{\tiny pre}}:&&\qquad h_-(z)=0,\qquad h_+(z)\propto z^{\Df-2-\ve}\, P^{(-\ve,\ve)}_{n}\left(\frac{2-z}z\right).
\eea
Here the superscript stands for `pre-functionals'. We call these objects pre-functionals, since we have not yet imposed on them the finiteness and swapping properties. With a suitable choice of normalization factor we have:
\begin{subequations}
\begin{align}
\tau_n^{\mbox{\tiny pre}}(\Delta_m)&=\delta_{nm}, &\tau_n^{\mbox{\tiny pre}}(\hD_m)&=0 \\
\widehat \tau_n^{\mbox{\tiny pre}}(\Delta_m)&=0,&\widehat \tau_n^{\mbox{\tiny pre}}(\hD_m)&=\delta_{nm}
\end{align}
\end{subequations}
with $\hD_n=1+\ve+n$ and $\Delta_m=2+2\ve+2n$ as before.
In particular, we see that as long as $\Df=\ve+1+k$ with $k\geq 0$ some integer, these functionals are dual to those blocks appearing in the crossing equation for the generalized free field solutions. 

In general we have more prefunctionals than we need, in the sense that some of these are not dual to any blocks appearing in the decomposition of the generalized free two point function. For $\Df=\ve+1+k$ with $k\geq 0$, we see that the bulk channel prefunctionals with $n=0,1,\ldots,k-1$ are not required. As for the boundary channel, we do not need the functionals with $n=0,1,\ldots k-1$ for Neumann boundary condition and in the Dirichlet case those with $n=0,1,\ldots,k$. These unwanted functionals can be put to good use to construct bona fide functionals (satisfying the swapping property) from the prefunctionals above. 

The construction is clearer to understand with an example. Let us set $\Df=\ve+1$. The prefunctionals $\widehat \tau_n$ have kernels $h_+(z)$ that do not falloff sufficiently fast at infinity to satisfy conditions \reef{eq:constraints}. Hence we must proceed with caution. There is a simple thing to do which is merely to consider linear combinations of prefunctionals that do falloff faster. One particular choice is to set
\bea
\widehat \tau_{n}=\widehat \tau_n^{\mbox{\tiny pre}}-c_n\widehat \tau_{0}^{\mbox{\tiny pre}}
\eea
for some suitable coefficient $c_n$, so as to  improve the falloff at infinity from the corresponding kernels from $1/z$ to $1/z^2$. It is clear many other choices of subtractions will do, but this is canonical in a sense, since recall that we can definitely afford to lose $\widehat \tau_0^{\text{\tiny pre}}$ for the Dirichlet boundary condition. For Neumann it looks like we are apparently missing a functional, but there is a good reason for this, as will be explained further below. 

After this subtraction procedure we have perfectly good boundary functionals. We can proceed in the same way for the bulk channel functionals, defining
\bea
\tau_n=\tau_n^{\mbox{\tiny pre}}-d_n \widehat \tau_0^{\mbox{\tiny pre}}
\eea
with $d_n$ chosen such that $h_+(z)-h_-(z)$ decays like $1/z^2$ at infinity. In this way we construct a set of functionals satisfying finiteness, swapping and the duality conditions:
\ba
\tau_n^{}(\Delta_m)&=\delta_{nm}, &\tau_n^{}(\hD_m)&=-d_n \delta_{m,0},&\qquad n,m&\geq 0 \\
\widehat \tau_n^{}(\Delta_m)&=0,&\widehat \tau_n^{}(\hD_m)&=\delta_{n,m}-c_n \delta_{m,0},&\qquad n\geq 1, m&\geq 0.
\ea
where now $\Delta_m=2\Df+2m$, $\hD_n=\Df+n$.
This procedure can be repeated for other values of $\Df=1+\ve+k$ in a similar manner. Everytime we increase $\Df$ by one unit the asymptotic behaviour of the prefunctionals is worse by an extra factor of $z$, but this is compensated by there being more prefunctionals to use. Not all of these are independent, and in the end the counting of degrees of freedom always works out that we end up with just enough functionals to obtain the duality relations written above.

\subsubsection{Shifted functionals and general $\Df$}

Let us go back to our simple example at $\Df=1+\ve$. In this case it is easy to check that for $n=1$ one gets:
\bea
\widehat \tau_1:\qquad h_+(z)\propto \frac{1}{z^2}.
\eea
In fact, one can check that for any $\Df=1+\ve+k$ one always finds this very same result. The reason for this is easy to understand. The Jacobi polynomials can be written as polynomials in $z^{-1}$, and so by taking linear combinations it must be that a particularly simple set of functionals satisfying finiteness and swapping can be defined as:
\bea
\boxed{
s\widehat {\tau}_n: \qquad h_-(z)=0,\qquad h_+(z)\propto \frac{1}{z^{1+n}},\qquad n \geq 1.
}
\eea
The letter 's' stands for ``simple'' or ''shifted'', since we see all kernels are simply determined from the $n=1$ case by shifting the exponent or multiplication by a power of $z^{-1}$.

The shifted functionals satisfy the duality conditions
\begin{subequations}
\label{eq:orthobdy}
\begin{align}
s\widehat{\tau}_n(\Delta_m)&=0& \mbox{for}&\quad m\geq 0, n\geq 1 ,\\
s\widehat \tau_n(\hD_m)&=\delta_{nm},& \mbox{for} &\quad m\geq n, \quad n\geq 1,
\end{align}
\end{subequations}
where $\hD_m=\Df+m=\ve+2+m$ and $\Delta_m=2\Df+2m$. 

The discussion goes through mutatis mutandis for the bulk channel, for which we can choose:
\begin{empheq}[box=\widefbox]{align}
s\tau_n:&& h_+(z)&=0,& h_{-}(z)&\propto \frac{1}{(1-z)^{1+n}},\qquad n\geq 1\\
s\tau_0:&& h_+(z)&\propto -\frac 1z,&h_{-}(z)&\propto \frac{1}{(1-z)}.
\end{empheq}
where for $s\tau_0$ we must choose proportionality factors that $h_+(z)-h_-(z)$ falls off as $1/z^2$ at infinity.
These functionals now satisfy
\begin{subequations}
\label{eq:orthobulk}
\begin{align}
s\tau_n(\Delta_m)&=\delta_{nm},& \mbox{for} &\quad m\geq n, \quad n\geq 0,\\
s\tau_n(\hD_m)&=-\delta_{m,0} e_m,& \mbox{for} &\quad m,n\geq 0,
\end{align}
\end{subequations}
for some contants $e_m$.

This particular functional basis has extremely simple kernels $h_{\pm}(z)$, and for this we paid the relatively small price that the orthonormality conditions are more involved. It is clear however that starting from this basis, for any $n$ a finite Gram-Schmidt decomposition can be made to go back to the orthonormal basis.

Now comes the crucial point. Although we have arrived at this basis via a somewhat circuitous route, the final functionals are extremely simple, and in particular all details related to the fact that we started with Jacobi polynomials have disappeared. This is not an accident, since the functionals constructed above will actually have the same properties for {\em any} $\Df$. 

To see this, consider for definiteness the action of the boundary functionals. The bulk case is analogous. When acting on bulk blocks we get
\bea
&&s\widehat \tau_n(\Delta)\propto \int_{1}^{\infty}\frac{\ud z}{z^{2+n}} \mbox{Im}\left[G_{\Delta}(z|\Df)\right]\nonumber\\
&=& \sin\left[\frac{\pi}2(\Delta-2\Df)\right]
\int_{1}^{\infty}\frac{\ud z}{z^{2+n}} (z-1)^{\frac{\Delta}2-\Df}\,\frac{{}_2F_1\left(\frac{\Delta}2,\frac{\Delta}2,\Delta-\varepsilon,\frac{z-1}z\right)}{z^{\frac{\Delta}2}}
\eea
Thanks to the oscillating factor we see the functional is automatically zero for $\Delta=2\Df+m$ independently of $\Df$, with $m\geq 0$.\footnote{For $\Delta\leq 2\Delta-2$ the integral diverges, and we must use instead expression \reef{eq:funcans} for the functional action.}. Now consider the action on the boundary blocks:
\bea
s\widehat \tau_n(\hD)\propto \int_{1}^{\infty} \frac{\ud z}{z^{2+n}} \mbox{Im}\left[ \hG_{\hD}(z|\Df)\right].
\eea
Starting from this expression we deform the contour of integration to wrap around the discontinuity of $\hG_\hD$ for negative $z$. We will see momentarily what are the conditions for this manipulation. If we do this we now get
\bea
s\widehat \tau_n(\hD)\propto \sin\left[\pi(\hD-\Df)\right] \int_{-\infty}^0 \frac{\ud z}{z^{2+n}} \frac{\hG_\hD\left(\mbox{$\frac{z}{z-1}$}|\Df\right)}{(1-z)^{\Df}}
\eea
We see that automatically the functional vanishes on boundary blocks with dimensions $\hD=\Df+m$, with $m$ integer. But this is true only down to $m=n$, since at this point the integral develops a divergence from the region near $z=0^-$. The divergence cancels the zero of the sine function, leading to a finite answer. For $\hD$ smaller than $\Df+1+n$ we can no longer trust the contour deformation and must go back to the original definition of the functional action. We conclude that $s\widehat \tau_n$ satisfies indeed the duality relations \reef{eq:orthobdy}. A similar argument goes through for the bulk functional basis. The messy details involving $\Df$ and $\ve$ only become important if we insist in having a fully orthonormal basis.

\subsection{Functional bootstrap equations}
To summarize, we have shown that an orthonormal basis of functionals $\tau_n, \widehat \tau_n$ exist with the properties:
\ba
\label{eq:orthotoy}
\tau_n(\Delta_m)&=\delta_{nm}, &\tau_n(\hD_m)&=-d_n \delta_{m,0},& \qquad n\geq 1,m&\geq 0 \\
\widehat \tau_n(\Delta_m)&=0,&\widehat \tau_n(\hD_m)&=\delta_{nm}-c_n \delta_{m,0},& \qquad n,m&\geq 0
\ea
with $\hD_m=\Df+m$ and $\Delta_m=2\Df+2m$. This basis can be obtained for any $\Df$ by starting with the shifted functionals $s\tau_n$ and $s\widehat \tau_n$ and performing a finite  Gram-Schmidt orthonormalization procedure.  Applying this functional basis to a generic crossing equation this gives a set of necessary conditions on the BCFT data:
\bea
\sum_{\hD} \mu^2_{\hD} \omega_n(\hD)=\sum_{\hD} (a\lambda)_\Delta \omega_n(\Delta),\qquad \omega=\tau,\widehat \tau,
\eea
which are {\em the functional bootstrap equations} associated to this particular functional basis. 

There is a different way to understand the origin of these equations. The orthonormality conditions suggest one may write
\begin{subequations}
\bea
G_{\Delta}(z)=\sum_{n=0}^{\infty}\left[\tau_n(\Delta) G_{2\Df+2n}(z)+\widehat \tau_n(\Delta) \hG_{1+\Df+n}(z)\right]\\
\hG_{\hD}(z)=\sum_{n=0}^{\infty}\left[\tau_n(\Delta) G_{2\Df+2n}(z)+\widehat \tau_n(\Delta) \hG_{1+\Df+n}(z)\right]
\eea
\end{subequations}
where we have ommitted the dependence of the blocks (and functionals) on $\Df$ for simplicity. These expressions can be thought of as decompositions of generic bulk and boundary blocks into a {\em basis}, the basis being constituted by bulk and boundary blocks with specific, generalized free, scaling dimensions. If these decompositions are actually true (we will not attempt to prove this here), then by plugging them into the original crossing equation we get
\ba
&\sum_{n=0}^{\infty}\left\{ \left[\sum_{\hD} \mu^2_{\hD} \tau_n(\hD)-\sum_{\hD} (a\lambda)_\Delta \tau_n(\Delta)\right]\,G_{2\Df+2n}(z)+\right. \\
 &\left.+\left[\sum_{\hD} \mu^2_{\hD} \widehat \tau_n(\hD)-\sum_{\hD} (a\lambda)_\Delta \widehat \tau_n(\Delta)\right]\,\widehat G_{1+\Df+n}(z)\right\}=0,
\ea
and the functional equations now follow demanding that the coefficient of each ``basis element'' is vanishing.

A closely related perspective on the same equations allow us to define a toy version of the Polyakov bootstrap. If we define the (toy) Polyakov blocks
\ba
P_{\Delta}(z):=G_{\Delta}(z)-\sum_{n=0}^{\infty}\tau_n(\Delta) G_{2\Df+2n}(z),\\
\widehat P_{\hD}(z):=\hG_{\hD}(z)-\sum_{n=0}^{\infty}\widehat \tau_n(\hD) \widehat G_{1+\Df+n}(z),
\ea
then it is not hard to show using the basis decompositions above that the functional bootstrap equations arise from demanding that correlators may be expressed as
\bea
\mathcal G(z)=\sum_{\hD} \mu^2_{\hD} \widehat P_{\hD}(z)+\sum_{\Delta} (a\lambda)_{\Delta} P_{\Delta}(z).
\eea 
Hence, this simple basis have allowed us to derive a toy version of the Polyakov bootstrap for BCFTs. Even though we have not attempted to, it is likely all the above statements can be made rigorous. In the next sections we will construct a different basis of functionals, leading to different functional bootstrap equations and different Polyakov blocks, but the basic ideas are as above. 

The functionals equations can be used for instance to reconstruct the generalized free solutions to crossing. Assuming the correct spectrum for each boundary condition, one can read off the BCFT data:
\ba
\mu^2_{\Df+1+2p}&=\widehat \tau_{1+2p}(0),& (a\lambda)_{2\Df+2p}&=-\tau_{p}(0),& &\text{Dirichlet}\\
\mu^2_{\Df+2p}&=\widehat \tau_{2p}(0)+c_{2p} \mu^2_{\Df},& (a\lambda)_{2\Df+2p}&=-\tau_{p}(0)-d_{p} \mu^2_{\Df},& &\text{Neumann}
\ea
Note we also have the constraints:
\ba
\widehat \tau_{2p}(0)&=0,&  &\qquad\text{Dirichlet}\\
\widehat \tau_{1+2p}(0)&=-c_{1+2p} \mu^2_{\Df},& &\qquad\text{Neumann}
\ea
which follow from applying these functionals to the Dirichlet and Neumann solutions respectively. From these relations we can reconstruct the whole OPE and BOE data for both solutions given $c_n, d_n$ which were fixed by demanding the correct asymptotics of the functional kernels.

It may seem puzzling that we are apparently missing a functional, namely $\widehat \tau_0$, which we had to remove to fix the behaviour of the kernels at $z=\infty$. However, this is precisely how it must be. Indeed, notice that by combining the above two solutions defining
\bea
\mathcal G(z)=2x-1+\frac{1}{(1-z)^\Df}=x \mathcal G^N(z)+(1-x) \mathcal G^{D}(z),
\eea
with $0\leq x\leq 1$, leads to a perfectly reasonable family of solutions to crossing which generically contain all operators in both GFF solutions. Had we been able to find a new functional such that the duality conditions would hold even for $\hat \Delta=\Delta_\phi$, i.e. with $c_n,d_n=0$. Then we could read off the OPE and BOE data simply by acting with the functionals to obtain
\bea
\mu_{\hat \Delta_n}^2=\widehat \tau_n(0),\qquad (a\lambda)_{\Delta_n}=\tau_n(0)
\eea
Since on the RHS of these equations we just have some fixed numbers, this would be in contradiction with the existence of the family of solutions described above. The fact that we have this degree of freedom in the choice of crossing solution which keeps the set $\hat \Delta_m=\Df+m$ and $\Delta_m=2\Df+2m$ fixed in the block decomposition, translates into the absence of a functional in the dual space. Indeed, $x$ controls the $O(1)$ behaviour of the correlator at $\infty$, and our functionals were carefully constructed such that they do not care about this overall constant.

We could now go on to investigate the consequences that these equations imply for generic solutions to crossing. However, it seems likely this will not be very useful. Consider for instance starting from the Dirichlet GFF solution and deforming it in a very slight manner. Unless this deformation is trivial like in the $x$- dependent family of correlators above, one should expect that at least the boundary spectrum should change non-trivially. While to zeroth order the equations are nicely diagonal in the spectrum thanks to orthonormality, allowing us to just read off the OPE and BOE data, the minute we expand in the spectrum the equations will become completely coupled. This is because $\partial_\Delta \omega_n(\Delta_m)\neq \delta_{nm}$ for these functionals. Clearly it would be nice to have a basis that would have these properties, since it would allow us to do perturbation theory in an efficient manner. Such a basis indeed exists, and we will construct it in the next section

Before we do so, we would just like to comment that in \cite{Bissi2018}, the authors were able to bootstrap the Wilson-Fisher fixed point to $O(\epsilon^2)$ by essentially using the basis of functionals constructed in this section, although they did not formulate their approach in this language.\footnote{To be precise they used our initial Jacobi polynomial basis and only for the bulk channel.}The reason this was possible is that the full set of boundary and bulk operators appear only at $O(\epsilon^2)$ and only a finite set of them appear with dimensions away from their free integer values. It is clear however that this approach is doomed to fail at order $O(\epsilon^3)$ where an infinite number of operators pick up anomalous dimensions. With the basis that we will construct next, there are no such limitations and it is possible in principle to go to arbitrarily high orders in perturbation theory.

\section{Bootstrap equations and BCFT basis}
\label{sec:functionals}
\subsection{Witten diagrams and functional equations}
\begin{figure}[H]

	\begin{center}
		\vskip 2pt
		\resizebox{450pt}{100pt}{\includegraphics{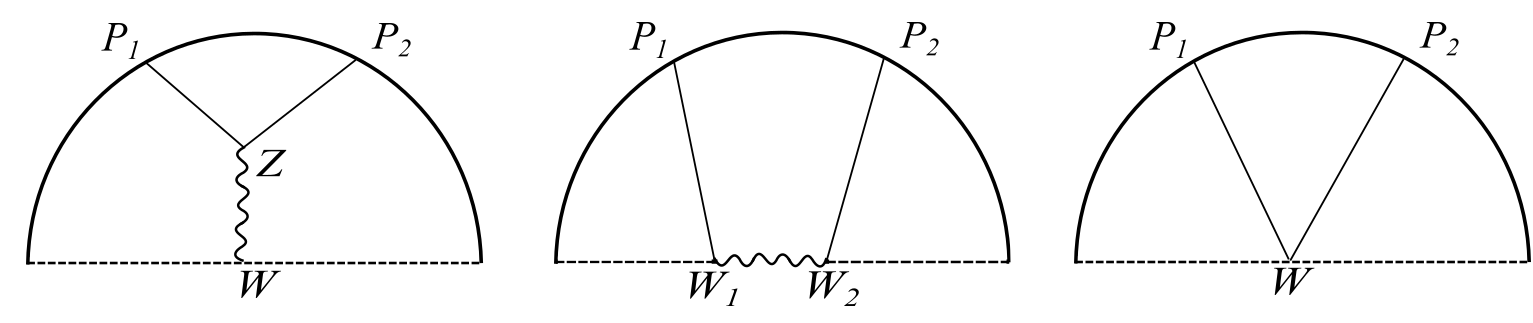}}
	\end{center}
	\caption{The bulk exchange, boundary exchange and contact Witten diagrams respectively. The semicircular area represents $AdS_{d+1}$ bulk in the presence of a brane, which is represented by the dashed line.  }
	\label{fig:bulkbdycon}
\end{figure}

As we have discussed in section \ref{sec:review}, the generalized free solutions to crossing can be understood as arising from the theory of a free massive scalar field in AdS$_{d+1}$ in the presence of a radially extended brane, on which Neumann or Dirichlet boundary conditions are imposed. In such a theory it is very natural to consider perturbations. We will focus on those perturbations which modify how the field interacts with the brane, while keeping the bulk theory fixed. At tree level there are three sets of perturbations that we can consider. We represent them diagramatically in figure \ref{fig:bulkbdycon}. They correspond to bulk and boundary exchange Witten diagrams, as well as contact interactions. Such diagrams have both bulk and boundary conformal block expansions. Crossing symmetry for each such diagram leads to the following equations:
\begin{subequations}
\label{eq:decompwitten}
\bea
G_{\Delta}(z)&=&\sum_{n=0}^{\infty}\left[a_n(\Delta)\,\hG_{\hD_n}(z)+b_n(\Delta)\,\partial_{\hD}\hG_{\hD_n}(z)+t_n(\Delta) G_{\Delta_n}(z)\right]\label{eq:bulkdecomp}\\
\hG_{\hD}(z)&=&\sum_{n=0}^{\infty}\left[\hat a_n(\hD)\,\hG_{\hD_n}(z)+\hat b_n(\hD)\,\partial_{\hD}\hG_{\hD_n}(z)+\hat t_n(\hD) G_{\Delta_n}(z)\right]\label{eq:bdydecomp}\\
0&=&\sum_{n=0}^{\infty}\left[d_n^{(k)}\,\hG_{\hD_n}(z)+e_n^{(k)}\,\partial_{\hD}\hG_{\hD_n}(z)+f_n^{(k)} G_{\Delta_n}(z)\right]
\eea
\end{subequations}
with $\Delta_n=2\Df+2n$ running over bulk double trace operators, $\hD_n=\frac{1-\nu}2+\Df+2n$ over boundary derivative operators and $k$ labels the contact interaction. We have omitted the dependence of the blocks on $\Df$ for brevity.

By analogy with the results of the last section, these expressions suggest that it may be possible to use the sets $G_{\hD_n}, \partial_{\hD} G_{\hD_n}$ and $G_{\Delta_n}$ as a {\em basis}. Our job in this section is then to construct functionals which are dual to this basis, i.e. $\alpha_n,\beta_n, \theta_n$ such that:
\begin{subequations}
\label{eq:normal}
\begin{align}
\alpha_n(\hD_m)&=\delta_{nm},&\partial_{\hD}\alpha_n(\hD_m)&=0,&\alpha_n(\Delta_m)&=0,\\
\beta_n(\hD_m)&=0,&\partial_\hD\beta_n(\hD_m)&=\delta_{nm},&\beta_n(\Delta_m)&=0,\\
\theta_n(\hD_m)&=0,&\partial_{\hD}\theta_n(\hD_m)&=0,&\theta_n(\Delta_m)&=\delta_{nm}
\end{align}
\end{subequations}
To be precise, there are really two different bases one can consider, depending on whether we choose functionals dual to the Neumann or Dirichlet solution. When necessary we will make this explicit by adding a superscript $+$, $-$ to the functionals for Neumann, Dirichlet respectively. We will see that in the latter case we will be able to indeed satisfy the above conditions, while for Neumann there will be a slight modification, as for the toy basis of the previous section, cf. equation~\reef{eq:orthotoy}.~\footnote{This is also very similar to what happens in the 1D bootstrap case, with our Dirichlet and Neumann cases being analogous to the fermionic and bosonic basis \cite{Mazac:2018ycv}.}

By acting with the functional bases on bulk and boundary blocks, one expects that decompositions of the form \reef{eq:decompwitten} should exist, with:
\bea
a_n(\Delta)\to \alpha_n(\Delta),\qquad b_n(\Delta)\to \beta_n(\Delta),\qquad t_n(\Delta)\to \theta_n(\Delta)\\
\hat a_n(\hD)\to \alpha_n(\hD),\qquad \hat b_n(\Delta)\to \beta_n(\hD),\qquad \hat t_n(\hD)\to \theta_n(\hD)
\eea
In other words, these functionals should allow us to bootstrap Witten diagrams in $AdS_{d+1}$. We will compute these Witten diagrams explicitly in the next section, and study the precise relation to the functional bases in section \ref{sec:polyakov}. 

Just like with the toy basis, we can act with these functionals on the crossing equation \reef{eq:crossingeq}, to obtain a set of {\em functional bootstrap equations}:
\begin{subequations}
\begin{empheq}[box=\widefbox]{align}
\label{eq:funcbootstrap}
\sum_{\hD} \mu_{\hD}^2 \alpha_n(\hD)&=\sum_{\Delta} (a\lambda)_{\Delta} \alpha_n(\Delta)\\
\sum_{\hD} \mu_{\hD}^2 \beta_n(\hD)&=\sum_{\Delta} (a\lambda)_{\Delta} \beta_n(\Delta)\\
\sum_{\hD} \mu_{\hD}^2 \theta_n(\hD)&=\sum_{\Delta} (a\lambda)_{\Delta} \theta_n(\Delta)
\end{empheq}
\end{subequations}
for all $n\in \mathbb N$.\footnote{Our definition of $\mathbb N$ is such that it is a monoid with the ordinary addition operation.}
The functional bootstrap equations are a set of fully non-perturbative constraints on the OPE and BOE data. Unlike the toy basis, the $\alpha,\beta,\theta$ basis now provide an ideal starting point for a perturbative expansion around the generalized free solution. 

To see this explicitly, consider starting with the free correlator and deforming it by adding a new operator with some small coefficient $g$. Expanding all relevant quantities,
\begin{subequations}
\bea
\mathcal G(z)&=&\mathcal G^{(0)}(z)+g\, \mathcal G^{(1)}(z)+O(g^2),\\
\Delta_n&=&2\Df+g \Delta_n^{(1)}+O(g^2),\qquad \hD_n=\frac{1-\nu}2+n+g \hD_n^{(1)}+O(g^2)\\
(a\lambda)_{\Delta_n}&=&\te^{(0)}+g\, \te^{(1)}+O(g^2),\qquad \mu^2_{\hD_n}=\mm_n^{(0)}+g \mm_n^{(1)}+O(g^2)
\eea
\end{subequations}
where $\te^{(0)}, \mm_n^{(0)}$ are the generalized free BCFT data given in \reef{eq:gffdata}, and 
\bea
\mathcal G^{(0)}(z)&=&G_0(z)+\sum_{n=0}^{\infty} \te_n^{(0)} G_{\Delta_n}(z)=\sum_{n=0}^{\infty} \mm_n^{(0)} \hG_{\hD_n}(z)\\
\mathcal G^{(1)}(z)&=& G_{\Delta}(z)+\sum_{n=0}^{\infty} \te_n^{(1)} G_{\D_n}(z)\nonumber\\
&=&\sum_{n=0}^{+\infty}\left[ \mm_n^{(1)} \hG_{\hD_n}(z)+\mm_n^{(0)} \hD_n^{(1)} \partial_{\hD} \hG_{\hD_n}(z)\right]
\eea
Acting with the functionals we read off:
\begin{subequations}
\begin{align}
\mm_n^{(0)}&=\alpha_n(0),&\mm_n^{(1)}&=\alpha_n(\Delta)\\
0&=\beta_n(0),& \mm_n^{(0)} \hD_n^{(1)}&=\beta_n(\Delta)\\
\te_n^{(0)}&=-\theta_n(0),&\te_n^{(1)}&=-\theta_n(\Delta)
\end{align}
\end{subequations}
In particular the functionals not only allow us to bootstrap the generalized free solution, but they also allow us to read off anomalous boundary operator dimensions under perturbations by using the $\beta$ functionals. It is not hard to see that this behaviour persists to arbitrary orders in perturbation theory, in other words, the $\alpha,\beta,\theta$ functional basis diagonalises perturbation theory around the generalized free solution as long as the bulk spectrum is kept fixed (or known). We will discuss a particular application in section \ref{sec:WF}, where we will bootstrap the Wilson-Fisher fixed point with a boundary with Neumann and Dirichlet boundary conditions.

After these motivating remarks, we will now turn to the actual construction of the functional basis. We begin by discussing general constraints on the functional kernels, before solving these constraints appropriately, first for the boundary channel functionals $\alpha,\beta$, and then the bulk functionals $\theta$.

\subsection{Constraining the kernels}

We start with the ansatz introduced in the previous section:
\bea
\omega[\mathcal F]:=\int_{-\infty}^{0} \ud z h_-(z) \mbox{Im}[\mathcal F(z)]+\int_{1}^{\infty} \ud z h_+(z) \mbox{Im}[\mathcal F(z)]
\eea
The orthonormality properties \reef{eq:normal} imply that we want to find functionals which have double zeros in the boundary channel. So, unlike for our warm-up basis, it is much less obvious how to choose appropriate kernels. However, one of the lessons of that construction is that it is useful to rotate contours so as to go to the region where discontinuities of the blocks are simple. 

Consider the action of a generic functional $\omega$ on the bulk blocks. Rotating contours we can write
\bea
\omega(\Delta)=-\int_{1}^{\infty}\ud z\, h_+(z) \mbox{Im}\left[G_{\Delta}(z|\Df)\right]+\int_0^1 \ud z\, \mbox{Im}\left[h_-(z)\right]\, G_{\Delta}(z|\Df)\nonumber\\
+\int_{1}^{\infty}\ud z\, \left\{\mbox{Im}\left[h_-(z)\right]\, \mbox{Re}\left[G_{\Delta}(z|\Df)\right]+\mbox{Re}\left[h_-(z)\right]\, \mbox{Im}\left[G_{\Delta}(z|\Df)\right]\right\}
\eea
We want functionals to annihilate the generalized free boson bulk spectrum. In the region $z>1$ we have that the real and imaginary parts of the bulk block are proportional to $\cos\left[\frac{\pi}2(\Delta-\Df)\right]$ and $\sin\left[\frac{\pi}2(\Delta-\Df)\right]$ respectively. Hence it is natural to keep only the latter terms in the expression above, which can be achieved by demanding
\bea
\mbox{Im}\left[h_-(z)\right]=0,\qquad \forall z\in \mathbb R.
\label{eq:bulkconds}
\eea
This is then one of our conditions on the kernels. For future reference, we note that in this case the functional action becomes
\bea
\omega(\Delta)=\sin\left[\frac{\pi}2(\Delta-2\Df)\right]\int_{1}^{\infty}\,\ud z\left[h_-(z)-h_+(z)\right] \,\frac{{}_2F_1\left(\frac{\Delta}2,\frac{\Delta}2,\Delta-\varepsilon,\frac{z-1}z\right)}{(z-1)^{\Df-\frac{\Delta}2}z^{\frac{\Delta}2}}.\label{eq:funcactionbulk}
\eea
This is valid only for $\Delta$ sufficiently large, depending on the singularity structure near $z=1$ of $h_-,h_+$. 

Next we look at the functional action on the boundary channel. In this case we want to do the contour rotation in a different way, swapping the roles of $h_+, h_-$ to obtain
\bea
\omega(\hD)=\int_{-\infty}^{0}\ud z\, h_-(z) \mbox{Im}\left[\hG_{\hD}(z|\Df)\right]-\int_0^1 \ud z\, \mbox{Im}\left[h_+(z)\right]\, \hG_{\hD}(z|\Df)\nonumber\\
-\int_{-\infty}^{0}\ud z\, \left\{\mbox{Im}\left[h_+(z)\right]\, \mbox{Re}\left[\hG_{\hD}(z|\Df)\right]+\mbox{Re}\left[h_+(z)\right]\, \mbox{Im}\left[\hG_{\hD}(z|\Df)\right]\right\}
\eea
We would like for the functional action to have double zeros. Since the imaginary and real part of the blocks oscillate as $\sin[\pi(\hD-\Df)]$ and $\cos[\pi(\hD-\Df)]$, we can achieve this by combining the non-oscillating and cosine pieces and setting the sine pieces to zero. That is to say, we demand:
\begin{subequations}
\label{eq:bdyconds}
\begin{align}
h_-(z)&=\mbox{Re}\left[h_+(z)\right],& 0&>z\\
\mbox{Im}\left[h_+(z)\right]&=-\nu \mbox{Im}\left[h_+\left(\mbox{$\frac{z}{z-1}$}\right)\right](1-z)^{\Df-2},& 0&<z<1.
\end{align}
\end{subequations}
which constitues a second set of constraints on the kernels.
Under these conditions we get
\bea
\omega(\hD)=-(1-\nu \cos[\pi(\hD-\Df)]) \int_0^1 \ud z\, \mbox{Im}\left[h_+(z)\right]\, \widehat{G}_{\hD}(z|\Df) \label{eq:funcactionbdy}
\eea
which will have generically double zeros for $\hD=\frac{1-\nu}2+\Df+2n$, again at least for $\hD$ sufficiently large that allows us to do the contour deformations. 

In the following we will find solutions to equations \reef{eq:bulkconds},\reef{eq:bdyconds} subject to  boundary conditions appropriate for $\alpha_n,\beta_n$ and $\theta_n$ functionals satisfying the orthonormality conditions \reef{eq:normal},  as well as the swapping and finiteness conditions \reef{eq:constraints}.

Before we do this, there are important simplifications to keep in mind. Firstly, since $h_-(z)$ is real for real $z$, these equations as well as the finiteness and boundary conditions \reef{eq:constraints} only care about the difference $h_+(z)-h_-(z)$. Hence by shifting $h_+(z)\to h_+(z)+h_-(z)$ we can get rid of $h_-(z)$ altogether and work only with $h_+(z)$.

Secondly, and more importantly, these equations are invariant under the shifts:
\bea
\label{eq:ladder}
h_+(z)\qquad & \rightarrow &\qquad  z^{-p}(1-z)^{-q}\,h_+(z),\nonumber \\
\Df\qquad & \rightarrow &\qquad \Df-2q-p,\\
\nu\qquad & \rightarrow &\qquad (-1)^p\nu,\nonumber
\eea
with $p,q \in \mathbb Z$. This means that if we know a solution with specified singular behaviour near $z=0,z=1$ it is easy to obtain all remaning solutions.

\subsection{Boundary functionals}
\subsubsection{General solution}
Boundary functionals are comprised of the $\alpha_n$ and $\beta_n$. The orthonormality conditions \reef{eq:normal} tell us the corresponding functional actions have to be zero on all double trace blocks in the bulk channel. In the boundary channel they must be zero for all $\hD_m$ larger than $\hD_n$. At that point there are two possibilities. $\beta_n$ functionals are still zero for $\hD=\hD_n$, but non-zero on the derivative of a boundary block. $\alpha_n$ functionals are non-zero on $\hD=\hD_n$. Given the functional actions \reef{eq:funcactionbdy} and \reef{eq:funcactionbulk} we conclude that we must demand the behaviours:
\begin{subequations}
\bea
\alpha_n,\beta_n:&& h_+(z)\overset{z\to 1^+}\sim (1-z)^{-1+\eta},\quad \eta>0\\
\alpha_n:&& \mbox{Im}[h_+(z)]\overset{z\to 0^+}\sim \frac {\log(z)}{z^{\frac {3-\nu}2+2n}}\\
\beta_n:&& \mbox{Im}[h_+(z)]\overset{z\to 0^+}\sim \frac 1{z^{\frac {3-\nu}2+2n}}
\eea
\end{subequations}
where the first condition guarantees that representation \reef{eq:funcactionbulk} is valid strictly above $\Delta=2\Df$.

To obtain these functionals, we consider the set of kernels:
\bea
h_{m,k}(z):=-\frac{4^{k-m}}{\pi^{\frac 32}}\, \frac{\Gamma(1+\Df+m)}{\Gamma(1+m-k)}\frac{\,{}_2\tilde F_1\left(k-m-\frac 12,\frac 32+m-k,\frac 12+\Df+k,\frac 1z\right)}{z^{\frac 32+k}}.\label{eq:bdykernels}
\eea
with $_2\tilde F_1$ the regularized hypergeometric function and $m,k\leq m$ two non-negative integers.
We claim that functionals defined by $h_-(z)=0, h_+(z)=h_{m,k}(z)$ form a complete set from which finite linear combinations can be taken to obtain orthonormal functionals. Firstly, it is straightforward to check that these kernels satisfy equations \reef{eq:bdyconds} as long as $\nu=(-1)^{m+1}$. Then: 
\begin{subequations}
\bea
h_{m,k}(z)&\overset{|z|\to \infty}{\sim}& O[|z|^{-\frac 32-k}]\\
\mbox{Im}[h_{m,k}(z)]&\overset{z\to 1^-}{\sim}& O[(1-z)^{\Df+k-1/2}]\\
h_{m,k}(z)&\overset{z\to 1^+}{\sim}& O(1),\qquad \Df>\frac 12 \bigvee k>0\\
h_{m,0}(z)&\overset{z\to 1^+}{\sim}& O[(z-1)^{\Df-1/2}],\qquad \Df<\frac 12, k=0.
\eea
\end{subequations}
The first two relations above guarantee that these provide good functionals satisfying finiteness and swapping. The last two tells us they are indeed zero on all bulk blocks. To see these kernels form a good basis for the $\alpha_n, \beta_n$, we examine the singular behaviour of their imaginary parts near $z=0$. We find:
\begin{align}
h_{0,0}&:& O(z^{-2})&& \rightarrow& \beta_0^-\\
h_{1,0},h_{1,1}&:& O(z^{-3}),O[\log(z) z^{-1}]&&\rightarrow &  \beta_1^+, \alpha_0^+\\
h_{2,0},h_{2,1},h_{2,2}&:& O(z^{-4}),O(z^{-2}),O(\log(z)z^{-2})&&\rightarrow& \beta_1^-,\beta^-_0,\alpha_0^-\\
\vdots&& \vdots&&\vdots\nonumber
\end{align}
Note that the correspondence is not direct. We find that for even $m$, kernels with $k\leq m$ contain $\beta_0^-,\ldots,\beta_{m/2}^-$ and for $k>m$ we get combinations of $\alpha_0^-,\ldots,\alpha_{m/2-1}^-$ with the same $\beta$ functionals. So, even $m$ contains all Dirichlet functionals. For odd $m$, kernels with $k\leq (m+1)/2$ contain combinations of $\beta_1^+,\ldots,\beta_{(m+1)/2}^+$, and for $k>(m+1)/2$ we get combinations of $\alpha_0^+,\ldots, \alpha_{(m-1)/2}^+$ with the same $\beta$ functionals. In particular we find

\begin{subequations}
\label{eq:lowfuncs}
\bea
\beta_0^-:&&\qquad h_+(z)=h_{0,0}(z)=\frac{\Gamma \left(\Delta _{\phi }+1\right)}{\pi^{\frac 32}}z^{-\frac 32} \, _2\tilde F_1\left(-\frac{1}{2},\frac{3}{2};\Delta _{\phi
   }+\frac{1}{2};\frac{1}{z}\right)\label{eq:b0m}\\
\beta_1^+:&&\qquad h_+(z)=h_{1,0}(z)=\frac{\Gamma \left(\Delta _{\phi }+2\right)}{4\pi^{\frac 32}}z^{-\frac 32} \, _2\tilde F_1\left(-\frac{3}{2},\frac{5}{2};\Delta _{\phi
   }+\frac{1}{2};\frac{1}{z}\right)\\
\alpha_0^-:&&\qquad h_+(z)=\frac{8}{(1+\Df)(2+\Df)}\left(h_{2,2}(z)-h_{2,0}(z)+n_{\Df} h_{0,0}(z)\right)\\
\alpha_0^+:&&\qquad h_+(z)=\frac{8}{\Df(1+\Df)}\left(h_{1,0}(z)+h_{1,1}(z)\right)
\eea
\end{subequations}
with
\bea
n_{\Df}=\frac{(\Df+1)(\Df+2)}{32}\,(1+8\log(2)-4\,H_{\Df})
\eea
and $H_n$ the harmonic number function. 

\subsubsection{Neumann contact term ambiguity}
There is a slight problem in the above. Although it is not clear from the notation, as we have defined it $\partial_{\hD}(\alpha_0^+)(\Df)\neq 0$ as required by \reef{eq:normal}, and it would only be by adding a suitable multiple of $\beta_0^+$ that we could make this vanish. But $\beta_0^+$ has not made an appearance in the set of kernels above. It should correspond to something like a kernel with $m=k=-1$ but this then does not have the correct asymptotic behaviour at infinity.

In fact, such a functional does not exist, and the reason is very similar to what happened for the easy basis in section \ref{sec:easy}. Here the reason why such a functional cannot exist, is that it would preclude the existence of a relevant contact interaction in AdS$_{d+1}$ between the bulk field and the brane, namely $\int_{AdS_{d}}\ud^d x\sqrt{g} \Phi^2(\vec x, x^\perp,y=0)$. This decays at large $z$ and is only possible for Neumann boundary conditions. Such a contact interaction  deforms the Neumann GFF solution while introducing  anomalous dimensions to boundary operators that decay sufficiently fast for large $\hD$.\footnote{This is entirely analogous to what happens in the 1d functional construction  when one attempts to construct a basis associated to the generalized free boson \cite{Mazac:2018ycv}.} It is easy to check that had we been able to construct a full orthonormal basis of Neumann functionals, this contact term would've been ruled out, since it is a modification of a solution to crossing that does not introduce new states. 

Hence, a fully orthonormal basis cannot exist for the Neumann case. Instead, the best we can do is to find a set of $\alpha_n,\beta_n$ with $\beta_0\equiv 0$, such that
\begin{subequations}
\label{eq:normalNeum}
\begin{align}
\alpha_n^+(\hD_m)&=\delta_{nm},&\partial_{\hD}\alpha_n^+(\hD_m)&=-\delta_{m0}d_n,&\alpha^+_n(\Delta_m)&=0,\\
\beta_n^+(\hD_m)&=0,&\partial_{\hD}\beta_n^+(\hD_m)&=\delta_{nm}-\delta_{m0}e_n,&\beta_n^+(\Delta_m)&=0,
\end{align}
\end{subequations}
with $e_0\equiv 1$. The parameters $d_n, e_n, f_n$ are functions $\Df$ and spacetime dimension and follow from the the construction of the functional kernels. They can be determined by explicit computation of the functional actions, as shown in appendix \ref{app:funcactions}. The constants $f_n$ will make their appearance in the next subsection when we construct bulk functionals.

Thanks to these modified orthonormality conditions, the functionals now allow for a homogeneous solution to crossing:
\bea
0&=&\sum_{n=0}^{\infty}\left[d_n\,\hG_{\hD_n}(z|\Df)+e_n\,\partial_{\hD}\hG_{\hD_n}(z|\Df)+f_n G_{\Delta_n}(z|\Df)\right]\label{eq:contactope}.
\eea
This is of course nothing but the contact interaction described above. This will be checked by direct computation of the Witten diagram in section \ref{sec:witten}.

To conclude this subsection, let us explain why we refer to this as an ``ambiguity'' in the choice of basis. The point is that, unlike for the Dirichlet case, for Neumann it is simply impossible to have a fully orthonormal set of basis functionals with the correct large $z$ behaviour, and once we have given up on full orthonormality, there is no good reason to hold relations \reef{eq:normalNeum} to be sacrosanct. Equivalently there is now no canonical choice of basis, unlike for the Dirichlet functionals. In particular in the above we have decided to set $\beta_0\equiv 0$, but it is clear we could have chosen differently, e.g. by shifting every functional by $\beta_k$ we could have set $\beta_k=0$ instead, or we could have even set some finite linear combination of functionals to vanish. It is in this sense that the choice of basis is ambiguous. Nevertheless, the choice above is particularly simple and we shall stick to it.

\subsubsection{Shifted functionals}

We will now show a different way of constructing the functional basis, by using the solution-generating shifts introduced in \reef{eq:ladder}.

Let us set:
\begin{subequations}
\bea
\beta_0^-:\qquad h_+(z)\equiv h_{\beta_0^-}(z|\Df),\\
\alpha_0^-:\qquad  h_+(z)\equiv h_{\alpha_0^-}(z|\Df),\\
\beta_1^+: \qquad h_+(z)\equiv h_{\beta_1^+}(z|\Df),\\
\alpha_0^+: \qquad  h_+(z)\equiv h_{\alpha_0^+}(z|\Df),
\eea
\end{subequations}
i.e. the kernels on the righthand side can be read off from \reef{eq:lowfuncs}. Then we can define
\bea
s\beta_n^-:\qquad h_+(z)=z^{-2n} h_{\beta_0^-}(z|\Df+2n)\\
s\alpha_n^-:\qquad h_+(z)=z^{-2n} h_{\alpha_0^-}(z|\Df+2n)
\eea
These functionals satisfy all the desired constraints and have the correct singular behaviour near $z=0$. By taking finite linear combinations of $\omega_n,\omega_{n-1},\ldots \omega_0$ with $\omega=s\alpha^-,s\beta^-$ these can be chosen to satisfy the orthonormality conditions \reef{eq:normal}. This completes the construction of the Dirichlet boundary functionals. As for the Neumann case, first we can check that indeed:
\bea
h_{\beta_1^+}(z|\Df)=z^{-1} h_{\beta_0^-}(z|\Df+1)
\eea
Then we can define
\bea
s\beta_n^+:\qquad h_+(z)=z^{1-2n} h_{\beta_0^-}(z|\Df-1+2n)\\
s\alpha_n^+:\qquad h_+(z)=z^{1-2n} h_{\alpha_0^-}(z|\Df-1+2n)
\eea
for $n\geq 1$. For $n=0$ this method fails, since we would break the required asymptotic condition near $z=\infty$. For $\alpha_0^+$ however there is a way out: we can form a suitable linear combination of $h_{\beta_0^-}$,$h_{\alpha_0^-}$ which falls off not as $z^{-3/2}$ as each individual kernel, but as $z^{-5/2}$. We can then set
\bea
\alpha_0^+:\qquad h^+(z)=z\left[h_{\alpha_0^-}(z|\Df-1)+A(\Df)h_{\beta_0^-}(z|\Df-1)\right].
\eea
The choice $A(\Df)=H_\Df-\log(4)$ does the trick, and we can check that the resulting kernel matches the one we obtained in \reef{eq:lowfuncs}. This procedure does not work for obtaining $\beta_0^+$, since we are not allowed to use $\alpha_0^-$, so we conclude again that such a functional cannot be constructed.

\subsection{Bulk functionals}
We now turn our attention to the bulk functionals $\theta_n$. The orthonormality conditions \reef{eq:normal} tell us the corresponding functional actions must have double zeros on all double trace blocks in the boundary channel. In the bulk channel they must be zero for all $\Delta_m=2\Df+2m$ larger than $\Delta_n$. Given the functional actions \reef{eq:funcactionbdy},\reef{eq:funcactionbulk} we conclude that we must demand the behaviours:
\begin{subequations}
\begin{align}
\theta_n:&&\mbox{Im}\left[h_{+}(z)\right]&\overset{z\to 0^+}{\sim} O[z^{-1-\frac{1-\nu}2+\eta}],&\eta&>0\\
\theta_n:&& h_+(z)&\overset{z\to 1^+}\sim (z-1)^{-1-n},&&\end{align}
\end{subequations}
where the first condition guarantees that representation \reef{eq:funcactionbdy} is valid all the way down to $\hD=\frac{1-\nu}2+\Df$, and in particular that the functional action is zero for all boundary operators. As may be expected from the discussion of the Neumann case in the previous subsection, we should expect that something should go wrong when $\nu=1$, and that's indeed what we will find.

After a somewhat circuitous route\footnote{This involved constructing particular solutions for various integer $\Df$, obtaining series expansions for large $z$, and finally guessing the general $\Df$ result.} we were led to the following simple result for the lowest Dirichlet bulk functional:
\begin{align}
\theta_0^-:&&
h_+(z)&=h_{\theta_0^-}(z|\Df):=
N_{\Df}\sqrt{z}\left[\frac{\left(\Df-\frac 12\right)}{2(z-1)}\,{}_2\tilde F_1\left(-\frac 12,\frac 32,\frac 12+\Df;\frac 1z\right)+\right.\nonumber\\
&&&\left.+{}_2\tilde F_1\left(-\frac 12,\frac 12,\Df-\frac 32;\frac 1z\right)-\,{}_2\tilde F_1\left(-\frac 12,-\frac 12,\Df-\frac 32;\frac 1z\right)\right].
\end{align}
with the normalization factor chosen such that $\theta_0^-(2\Df)=1$,
\bea
N_\Df=\frac{4\Gamma(\Df-1)\Gamma(\Df+1)}{\pi\Gamma\left(\frac 12+\Df\right)}.
\eea
The functional kernel is a combination of two pieces which separately satisfy \reef{eq:bdyconds}, namely the top and bottom lines in the above. In particular the first piece is nothing but $\frac{z^2}{z-1}$ times the boundary functional with kernel $h_{0,0}(z)$ c.f. \reef{eq:bdykernels} (which thanks to \reef{eq:ladder} is again a solution). The specific combination above has the properties
\begin{subequations}
\bea
h_+(z)&\overset{z\to \infty}{\sim} &O(z^{-\frac 32})\\
\mbox{Im}[h_+(z)]&\overset{z\to 1^-}{\sim} &O[(1-z)^{\Df-\frac 12}]\\
h_+(z)&\overset{z\to 1^+}{\sim} &\frac 2\pi \frac{1}{z-1}\\
\mbox{Im}[h_+(z)]&\overset{z\to 0^+}{\sim} & O[\log(z)]
\eea
\end{subequations}
The first two conditions guarantee finiteness and swapping, and the bottom two characterize it as $\theta^-_0$. From this expression it is now easy to obtain all other Dirichlet functionals and (almost) all Neumann ones, by using the shift property \reef{eq:ladder}:
\bea
s\theta^-_n:&&\qquad h_+(z)=\frac{1}{(z-1)^{n}} h_{\theta_0^-}(z|\Df+2n),\qquad n\geq 0\\
s\theta^+_n:&&\qquad h_+(z)=\frac{z}{(z-1)^{n}} h_{\theta_0^-}(z|\Df-1+2n),\qquad n\geq 1
\eea
We cannot allow $n=0$ in the last equation since this would destroy the behaviour near $z=\infty$. As discussed in the previous subsection this was expected: for Neumann boundary conditions we cannot build fully orthonormal functionals since this would rule out the contact term interaction. This also give us the clue on how to obtain the $ \theta_0^+$: we must combine it with a boundary functional. This will mean that $\theta_0^+$ will no longer have double zeros on all boundary double trace blocks, and upon the Gram-Schmidt decomposition to orthonormalise the $\tilde \theta_n^+$ this will be the case for all functionals. A suitable linear combination is
\bea
\theta_0^+:\qquad h_+(z)=z\,\left[ h_{\theta_0^-}(z|\Df-1)-\frac{\sqrt{\pi}\Gamma(\Df)}{\Gamma\left(\Df+\frac 12\right)}, h_{\beta_0^-}(z|\Df-1)\right]
\eea
where $h_{\beta_0^-}(z|\Df)$ is a boundary functional kernel, cf. \reef{eq:b0m}, where we've added an extra label to show the dependence on $\Df$. The shift and multiplication by $z$ was necessary to obtain a Neumann type boundary kernel. The linear combination above falls off at infinity as $z^{-3/2}$ unlike each individual term. One can check all other requirements on the functional are satisfied, except that now near $z=0$ we have $\mbox{Im}[h_+(z)]\sim O(1/z)$. Hence this functional will have a first order zero at $\hD=\Df$. 

The conclusion is that, after orthonormalisation, the best we can do for the Neumann basis is:
\begin{align}
\theta_n(\hD_m)&=0,& \partial_{\hD}\theta_n(\hD_m)&=-\delta_{m0}f_n,& \theta_n(\Delta_m)=\delta_{nm}\label{eq:normalNeum2}
\end{align}
for $m,n\geq 0$. The constants $f_n$ are related to a contact diagram and given in appendix \ref{contdetails}.

\subsection{Functional actions}

Given the functional kernels $h_+(z)$, we are interested in determining the functional actions. We will now summarize how this is done, leaving specific results to appendix \ref{app:funcactions}.

For the purposes of computing the $\alpha$ and $\beta$ functional actions, we will use 
\bea
\omega_{m,k}:\quad h_{m,k}(z)=-\frac{\Gamma(1+\Df+m)}{\Gamma(1+m-k)}\frac{\,{}_2\tilde F_1\left(k-m-\frac 12,\frac 32+m-k,\frac 12+\Df+k,\frac 1z\right)}{4^{m-k} \pi^{\frac 32}z^{\frac 32+k}}
\eea
For the bulk functionals we will compute the shifted functionals from the kernels
\ba
s\theta^+_n:&&\qquad h_+(z)&=\frac{z}{(1-z)^n} h_{\theta_0^-}(z|\Df-1+2n)\\
s\theta^-_n:&&\qquad h_+(z)&=\frac{1}{(1-z)^n} h_{\theta_0^-}(z|\Df+2n)
\ea
with
\begin{align}
h_{\theta_0^-}(z|\Df)&:=
N_{\Df}\sqrt{z}\left[\frac{\left(\Df-\frac 12\right)}{2(1-z)}\,{}_2\tilde F_1\left(-\frac 12,\frac 32,\frac 12+\Df,\frac 1z\right)-\right.\nonumber\\
&\left.-{}_2\tilde F_1\left(-\frac 12,\frac 12,\Df-\frac 32,\frac 1z\right)+\,{}_2\tilde F_1\left(-\frac 12,-\frac 12,\Df-\frac 32,\frac 1z\right)\right].
\end{align}
and the normalization factor:
\bea
N_\Df=\frac{4\Gamma(\Df-1)\Gamma(\Df+1)}{\pi\Gamma\left(\frac 12+\Df\right)}.
\eea
With our choice of kernels we have in particular
\bea
s\theta^{\pm}_n(2\Df+2n)=1.
\eea
To evaluate the functional actions we have proceeded as follows. For the $\omega_{m,k}$ functionals we used the representations
\ba
\omega_{m,k}(\Delta)&=\int_{1}^{\infty}\ud z\, h_{m,k}(z) \mbox{Im}\left[ G_{\Delta}(z|\Df)\right]\\
\omega_{m,k}(\hD)&=\int_{1}^{\infty}\ud z \,h_{m,k}(z) \mbox{Im}\left[ \hG_{\hD}(z|\Df)\right]\\
\ea
The discontinuities of the blocks are given in terms of Gaussian hypergeometric functions, and so is the kernel $h_{m,k}$. We have evaluated these expressions by replacing the hypergeometric functions by their series representations and commuting the integration with the double series.

For the shifted bulk functionals we found it necessary to use different representations depending on which action was considered. For the action on bulk blocks we did as for the $\omega_{n,k}$. However, on boundary blocks it was necessary to use instead
\bea
s\theta_n(\hD)=-\left(1-\nu \cos\left[\pi(\hD-\Df)\right]\right)\int_0^1\ud z\, \mbox{Im}\left[h_{s\theta_n}(z)\right] G_{\hD}(z|\Df).
\eea
In both cases the computation of the functional action proceeds as for $\omega_{m,k}$. However, for the bulk functionals the final answer leaves one of the series unperformed.

To conclude note that given the functional actions above, one can go on to obtain the orthonormal functionals by performing a finite Gram-Schmidt orthonormalization, and in some cases we found it possible then to go on to guess the general orthonormal answer.

\section{Witten diagrams for BCFT}
\label{sec:witten}

\subsection{Setup}
We will now explicitly compute the Witten diagrams which motivated the functional basis constructed in the previous section. The Witten diagrams for external scalars with interface boundary conditions, were constructed in \cite{RastelliMellin} (see also \cite{Goncalves:2018fwx} for the defect case). For Neumann and Dirichlet boundary conditions, we will use similar techniques in our computations.
We consider a free scalar field propagating in AdS$_{d+1}$ in the presence of a brane extended along the radial direction. In the Poincar\'e patch the action is given by
\bea
S=\int_{x^\perp,y\geq 0} \frac{\ud y}{y^{d+1}}\ud^{d-1} x\, \ud x^{\perp} \left[\frac 12 \nabla \Phi\cdot\nabla \Phi+\frac 12 m^2 \Phi^2\right]
\eea
with $m^2=\Df(\Df-d)$ and boundary conditions:
\ba
\Phi(\vec x,x^\perp=0,y)&=0,& \qquad \nu&=-1& &\qquad \mbox{(Dirichlet)}\\
\partial_{\perp}\Phi(\vec x,x^\perp=0,y)&=0,& \qquad \nu&=1& &\qquad \mbox{(Neumann)}
\ea
The bulk field $\Phi$ is dual to the boundary operator $\phi$, which will satisfy the same boundary condition at $x^\perp=0$:
\ba
\Phi(\vec x,x^\perp,y)&\overset{y\to 0}{\sim} y^{\Df} \phi(\vec x,x^\perp)& &&&\\
\phi(\vec x,x^\perp=0)&=0,& \qquad \nu&=-1& &\qquad \mbox{(Dirichlet)}\\
\partial_\perp\phi(\vec x,x^\perp=0)&=0,& \qquad \nu&=1& &\qquad \mbox{(Neumann)}
\ea

In the absence of the brane, the bulk-to-boundary and bulk-to-bulk propagators would be\footnote{Up to normalization factors which will be irrelevant for us.} 
\bea\label{bulkbdyprop}
G_{B\partial}(P,X |\Df)&=&\frac{1}{(-2P\cdot X)^{\Df}}\\
G_{BB}(X_1,X_2|\Df)&=&\left(-\frac{1}\zeta\right)^{\Df}{}_2F_1\left(\Df,\Df-\frac{d-1}2,2\Df-d+1,-\frac{4}\zeta\right)
\eea
with $\zeta=\frac{(X_1-X_2)^2}4$. For economy of space we've introduced the embedding formalism - for a review and more details see \cite{Weinberg:2012mz,Penedones:2007ns,Paulos:2011ie}. In practice, we just need to know that $P^M, X^M$ are $d+2$-dimensional vectors satisfying $X^2=-1, P^2=0$, with
\bea
X^M&=&\left\{X^+,X^-,X^i,X^\perp\right\}=\frac{1}{y}\left\{1,|\vec x|^2+(x^\perp)^2+y^2,x^i,x^\perp\right\}\\
P^M&=&\left\{P^+,P^-,P^i,P^\perp\right\}=\left\{1,|\vec x|^2+(x^\perp)^2,x^i,x^\perp\right\}
\eea
with $i=1,\ldots, d-1$. This gives for instance 
\bea
-2 P_1\cdot P_2&=&|\vec x_{12}|^2+(x_1^\perp-x_2^\perp)^2\\
-2P_1\cdot X_2&=&\frac{1}{y_2}\left[|\vec x_{12}|^2+(x_1^\perp-x_2^\perp)^2+y_2^2\right]\\
\zeta&=&\frac{|\vec x_{12}|^2+(x_1^\perp-x_2^\perp)^2+(y_1-y_2)^2}{4y_1 y_2}.
\eea
In the presence of the brane, the propagators get modified. Since this is a free theory, it is straightforward to work out the correct propagators via the method of images. The result is that:
\begin{subequations}
\label{eq:propagators}
\bea
\label{bulkbulknu}G^{\nu}_{BB}(X_1,X_2|\Df)&=& G_{BB}(X_1,X_2|\Df)+\nu\, G_{BB}(X_{1r},X_2|\Df)\,,\\
G_{B\partial}^\nu(P,X|\Df)&=& G_{B\partial}(P,X|\Df)+\nu\,G_{B\partial}(P_{r},X|\Df)\nonumber \\
 &=& G_{B\partial}(P,X|\Df)+\nu\, G_{B\partial}(P,X_r|\Df)\,,
\eea
\end{subequations}
where a subscript 'r' indicates a reflection $x^\perp\to -x^\perp$. 

Below we will consider three kinds of Witten diagram. These correspond to exchanges of fields propagating in the AdS$_{d+1}$ bulk (bulk exchange), fields which propagate on the AdS$_d$ brane (boundary exchange) and finally we will also consider the simplest possible contact interaction with the brane. These diagrams will compute corrections to a BCFT$_d$ two point function $\langle \phi_1(\vec x_1,x_1^\perp)\phi_2(\vec x_2,x_2^\perp)\rangle$ of fields with dimensions $\Delta_1, \Delta_2$, although we will eventually be interested in setting $\phi_1=\phi_2=\phi$ and $\Delta_1=\Delta_2=\Df$. We will write for the two point functions:
\bea
W(P_1,P_2)=\frac{\mathcal W(z)}{(x_1^\perp)^{\Delta_1}(x_2^\perp)^{\Delta_2}},
\eea
where $z$ is the cross-ratio we've been using so far, and which can be written in terms of $P_1$ and $P_2$ explicitly as:
\bea
z=-\frac{2 P_1^{\perp} P_2^{\perp}}{P_1\cdot P_{2r}}.
\eea
We will stick to the general rule that hatted quantities relate to defect and unhatted to bulk. Hence Witten diagrams corresponding to boundary (bulk) exchanges will be represented by hatted (unhatted) quantities.

\subsection{Boundary exchange}\label{bdysectn}
The boundary exchange Witten diagram is represented in figure \ref{fig:bdywitten}. The diagram is then computed for the two different boundary conditions, as follows:

\begin{figure}[H]

	\begin{center}
		\vskip 2pt
		\resizebox{200pt}{120pt}{\includegraphics{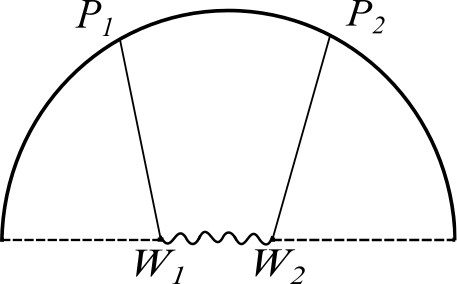}}
	\end{center}
	\caption{ The boundary exchange Witten diagram.}
	\label{fig:bdywitten}
\end{figure}

Let us put Neumann boundary conditions on the $AdS_{d+1}$ scalars $\Phi_1$ and $\Phi_2$, which are the  external legs of the diagram. We assume the vertices $\Phi_i \hat \Psi_{\hD}$, with $i=1,2$, on the $AdS_d$ defect. Here $\hat\Psi_{\hat\Delta}$ is a freely propagating scalar field on $AdS_d$, of dimension $\hat \Delta$. The boundary-bulk propagators end on the $AdS_d$ brane, where they become the same as ordinary AdS$_{d+1}$ bulk to boundary propagators (up to a factor of 2).  The diagram above then reduces to the same boundary exchange Witten diagram already evaluated in \cite{RastelliMellin}\,,
 \be\label{Wbdygen}
\widehat{W}^{+}(P_1,P_2)=\int_{AdS_{d}}dW_1dW_2 \  G_{B\partial}^{+}(P_1,W_1|\Delta_1)G_{B\partial}^{+}(P_2,W_2|\Delta_2)G_{BB}(W_1,W_2|\hD)\,.
\ee
The integration runs over the AdS$_d$ brane, and the measure is simply the ordinary one on AdS$_d$. The $G_{BB}(W_1,W_2|\hD)$ is an ordinary bulk-to-bulk propagator in $AdS_d$.

For Dirichlet boundary conditions, the vertex is instead $\partial_\perp \Phi_i \hat \Psi_{\hat\Delta}$ on the $AdS_{d}$ brane. The diagram has the following form,
 \be\label{WbdyDiri}
\widehat{W}^{-}(P_1,P_2)=\int_{AdS_{d}}dW_1dW_2 \  \partial_{\perp}^{W_1}G_{B\partial}^{-}(P_1,W_1|\Delta_1)\partial_{\perp}^{W_2}G_{B\partial}^{-}(P_2,W_2|\Delta_2)G_{BB}(W_1,W_2|\hD)\,.
\ee
Here we denote $\partial_\perp^W G(P,W)=\lim_{X \to W} \partial_{x^{\perp}} G(P,X)$. 
Near the brane we have:
\be\label{Dirichletprop}
G^-_{B\partial}(P,X) \overset{X\to W+O(x^\perp)}{\rightarrow}  \frac{x^\perp}{(-2P.W)}G^+_{B\partial}(P,W)\,.
\ee
Thanks to the derivatives in the integral \eqref{WbdyDiri}, the integral then becomes similar to the Neumann case, with boundary-bulk propagators whose external scalar dimensions have effectively increased by one. 

For unequal external scalar fields $\Phi_1, \Phi_2$, and Neumann boundary conditions the result is given as follows:
\be\label{WbdyNm}
\widehat{\mathcal W}^+_{\hD}(z)=\mathcal N^N_{\mbox{\tiny bdy}}(\hD)\int _{-i \infty }^{i \infty }d\tau \left[\frac{z}{4(1-z)}\right] ^{\tau }\frac{\Gamma (\tau ) \Gamma \left(h'-\tau \right)}{\Gamma \left(2 h'-2 \tau \right)}\int _{-i \infty }^{i \infty }dc\frac{\hat f(c,\tau ) \hat f(-c,\tau )}{(\hD -h')^2-c^2}\,,
\ee
where $h'=(d-1)/2$ \ and,
\be
\hat f(c,\tau)=\frac{\Gamma\left(\frac{c+\Delta_1-h'}{2}\right) \Gamma\left(\frac{c+\Delta_2-h'}{2}\right) \Gamma \left(c+h'-\tau \right)}{2 \Gamma (c)}\,.
\ee
The normalization factor is chosen such that when decomposing into boundary blocks we find:
\begin{align}\label{bdydeco2}
{\scalebox{0.9}{ $z^{-\frac{\D_1+\D_2}{2}}\widehat{\mathcal W}^+_{\hD}(z)$}}&={\scalebox{0.9}{ $\hG_{\hD}(z|\D_1,\D_2)$}}- \sum_{n=0}^{\infty}{\scalebox{0.8}{ $\bigg[\hat a_n^+(\Delta_1,\Delta_2,\hD) \hG_{\Delta_1+2n}(z|\D_1,\D_2)+ \hat a_n^+(\Delta_2,\Delta_1,\hD) \hG_{\Delta_2+2n}(z|\D_1,\D_2)\bigg] $}}\,.
\end{align}
Setting $\Delta_1=\Df+\frac{\Delta_{12}}{2},\Delta_2=\Df-\frac{\Delta_{12}}2$ and sending $\Delta_{12}\to 0$ we find
\bea
\hat a_n^+(\Delta_1,\Delta_2,\hD)\overset{\Delta_{12}\to 0}\sim \frac{\hat b_n^+(\Delta)}{\Delta_{12}}+\frac 12\hat a_n^+(\Delta)+O(\Delta_{12}).
\eea
We have ommitted the dependence of $\hat b_n^+, \hat a_n^+$ on $\Delta_1=\Delta_2=\Df$ for simplicity. The boundary block expansion becomes:
\bea
z^{-\D_\phi}\widehat{\mathcal W}^+_{\hD}(z)=\hG_{\hD}(z|\Df)-\sum_{n=0}^{\infty}\bigg[\hat a_n^+(\hD) \hG_{\Df+2n}(z|\Df)+ \hat b_n^+(\hD) \partial_{\hat{\Delta}} \hG_{\Df+2n}(z|\Df)\bigg]\,.
\eea
As for the bulk channel we get instead
\bea\label{bdydeco1}
\widehat{\mathcal W}^+_{\hD}(z)&=&z^{\frac{\Delta_1+\Delta_2}2}\sum_n \hat t_n^N(\Delta_1,\Delta_2,\hD)G_{\Delta_1+\Delta_2+2n}(z|\D_1,\D_2)\,.
\nonumber \\
&\overset{\Delta_{12}=0}=&z^{\Df}\sum_n \hat t_n^+(\hD)G_{2\Df+2n}(z|\Df)\,.
\eea

The Dirichlet Witten diagram $\widehat{\mathcal W}^-_{\hD}$ is obtained from $\widehat{\mathcal W}^+_{\hD}$, by shifting the  dimensions $\D_1,\D_2 \to \D_1+1,\D_2+1$\,. The corresponding boundary block coefficients can be directly obtained this way. The bulk block coefficients can be obtained be inspecting the expansion around $z\to 1$, where the expansion coefficients can be obtained from the above shifts. Explicit expressions for all coefficients appearing here are relegated to Appendix \ref{bdyExcdetails}\,.

\subsection{Bulk exchange}
\begin{figure}[h]
	\begin{center}
		\vskip 2pt
		\resizebox{400pt}{100pt}{\includegraphics{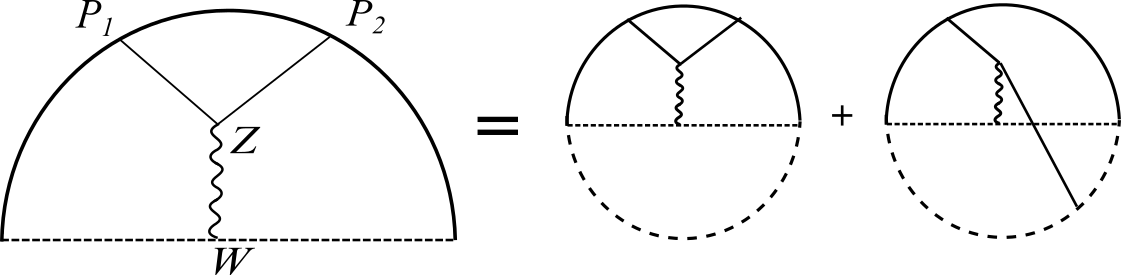}}
	\end{center}
	\caption{ The bulk exchange Witten diagrams can be written as a sum over two diagrams on full $AdS_{d+1}$. In the figure we show the Neumann case, Dirichlet would correspond to a minus sign on the RHS.}
	\label{fig:bulkexc}
\end{figure}

The bulk exchange Witten diagram is shown in figure \ref{fig:bulkexc}. For this diagram, we assume a vertex $\Phi_1\Phi_2\Psi_{\Delta}$ in the bulk, and a linear vertex $\Psi_{\Delta}$ on the brane. The $\Psi_\Delta$ is a freely propagating scalar operator in $AdS_{d+1}$ that gets exchanged. We put Neumann/Dirichlet boundary conditions on $\Phi_1$ and $\Phi_2$, but on $\Psi_\Delta$ we always put Neumann boundary conditions\footnote{We could have chosen a different boundary condition at the cost of modifying the vertex on the brane.}. The diagram is then given by:
\be
W_\Delta^\nu(P_1,P_2)=\int_{\frac{AdS_{d+1}}{2}}dX\int_{AdS_{d}}dW \  G_{B\partial}^{\nu}(P_1,X|\D_1)G_{B\partial}^{\nu}(P_2,X|\D_2)G_{BB}(X,W|\Delta)\,.
\ee
The integral label $\frac{AdS_{d+1}}{2}$ indicates that the integration is restricted to $x^{\perp}\geq 0$. Note that the bulk-to-bulk propagator is the ordinary one. This is easy to see since irrespective of the boundary conditions on $\Phi_1$ and $\Phi_2$, the exchange operator always has Neumann boundary condition. So the propagator $G^\nu_{BB}(X_1,X_2|\Delta)$ from \eqref{bulkbulknu} reduces to (twice of) $G_{BB}(X_1,X_2|\Delta)$ when one of the points  $X_1$ and $X_2$ lies on the $AdS_d$ brane. 

Using the propagators \reef{eq:propagators} and rearranging we can get:
\bea
\label{2intgrl}
W_\Delta^\nu(P_1,P_2)=\int_{AdS_{d+1}}dX\int_{AdS_{d}}dW \Big[ G_{B\partial}(P_1,X|\D_1)G_{B\partial}(P_2,X|\D_2)G_{BB}(X,W|\D)+\nonumber\\
+\nu\, G_{B\partial}(P_{1r},X|\D_1)G_{B\partial}(P_2,X|\D_2)G_{BB}(X,W|\D)\Big]\,.
\eea
These integrals can be computed using standard tools summarized nicely in \cite{RastelliMellin}. The first integral is straightforward, while the second is more involved. We give the details of this computation in Appendix  \ref{bulkdetails}\,. The final expression is given by:
{\small
	\begin{align}\label{Wbulk}
	&\mathcal W_\Delta^{\nu}(z) =\mathcal N_{\text{bulk}}^{\nu}(\Delta)  \int_{-i\infty}^{+i\infty} \frac{\ud\tau}{2\pi i}\, \left[\frac{4(1-z)}{z}\right]^{\tau-\frac{\Delta_1+\Delta_2}2}\, \frac{\G(\tau)\G(\tau-\frac{\D_{12}}{2})\G(\tau-\frac{\D_{21}}{2})}{\G(2\tau)\G(\frac{1}{2}-\tau)}\nonumber\\
&\times \int_{-i\infty}^{+i\infty} \frac{\ud c}{2\pi i}	
	\frac{f(c,\tau)f(-c,\tau)}{((\D-h)^2-c^2)}\left[1+\nu\,  \frac{{}_2 F_1\big( \frac{1-h+c}{2},\frac{1-h-c}{2},\frac{1}{2}-\tau ,z\big)}{(1-z)^{\frac{d-\Delta_1-\Delta_2-1}2}}\right]\,,
	\end{align}
}
where $h=d/2$ and,
\be
f(c,\tau)=\frac{\G(\frac{\D_1+\D_2-h+c}{2})\G(\frac{1+c-h}{2})\G(\frac{h+c}{2}-\tau)}{2\G(c)}\,.
\ee
The above expression\footnote{Note the presence of the extra hypergeometric ${}_2F_1$ function, which was not there for the interface case considered in \cite{RastelliMellin}. It is needed to implement the boundary condition. } has the following boundary channel decomposition (as  $z\to 0$),
\be\label{Wbbulkdeco1}
\mathcal W^{\nu}_\Delta(z)= z^{\frac{\D_1+\D_2}{2}}\sum_{n=0}^{\infty}{\scalebox{0.9}{ $\bigg[a_n^\nu(\D_1,\D_2,\D) \hG_{\Delta_1+2n+\frac{1-\nu}{2}}(z|\D_1,\D_2)+a_n^\nu(\D_2,\D_1,\D) \hG_{\Delta_2+2n+\frac{1-\nu}{2}}(z|\D_1,\D_2)\bigg]$}}\,.
\ee
The fact that we only get e.g. $\Delta_1$ plus an even integer depends crucially on the combination of terms in \reef{Wbulk}. As before, when we approach $\Delta_1\to \Delta_\phi+\frac{\D_{12}}2$ and $\Delta_2\to\Delta_\phi-\frac{\D_{12}}2$ we find
\bea\label{blockdeco1}
a_n^{\nu}(\Delta_1,\Delta_2,\D)\overset{\Delta_{12}\to 0}=\frac{ b_n^{\nu}(\Delta)}{\Delta_{12}}+\frac 12 a_n^{\nu}(\Delta)+O(\Delta_{12}),
\eea
and the boundary block expansion becomes:
\bea\label{blockdeco2}
\mathcal W^{\nu}_{\D}(z)=z^{\D_\phi}\sum_{n=0}^{\infty}\bigg[a_n^{\nu}(\D) \hG_{\Df+2n+\frac{1-\nu}{2}}(z|\Df)+ b_n^{\nu}(\D) \partial_{\hat{\Delta}} \hG_{\Df+2n+\frac{1-\nu}{2}}(z|\Df)\bigg]\,.
\eea
Finally, in the bulk channel we find
\bea
\mathcal W^{\nu}_{\Delta}(z)&=&z^{\frac{\Delta_1+\Delta_2}2}\left[G_{\Delta}^{\Delta_{12}}(z|\D_1,\D_2)-\sum_n t_n^\nu(\Delta_1,\Delta_2,\Delta)G_{\Delta_1+\Delta_2+2n}^{\Delta_{12}}(z|\D_1,\D_2)\right]
\nonumber \\
&\overset{\Delta_{12}=0}=&z^{\Df}\left[G_{\Delta}(z|\Df)-\sum_n t_n^\nu (\Delta)G_{2\Df+2n}(z|\Df)\right]\,.
\eea
All expansion coefficients are given later in Appendix \ref{bulkWittdetails}.

\subsection{Contact Diagrams}
The final tree level Witten diagrams we will consider are contact diagrams. They can be represented by two $AdS_{d+1}$ boundary-bulk propagators meeting at the $AdS_{d}$ boundary where they interact via a quadratic vertex. There are several kinds of such diagrams depending on the choice of this vertex. Here we will consider the simplest possible cases. 

With Neumann boundary condition, the simplest such vertex is $\Phi_1\Phi_2$ vertex on the $AdS_{d}$ brane. We have: 
\be\label{WContgen}
W_{\text{C}}^+=\int_{AdS_{d}}dW \  G_{B\partial}^{+}(P_1,W|\D_1)G_{B\partial}^{+}(P_2,W|\D_2)\,.
\ee%
\begin{figure}[H]
	\begin{center}
		\vskip 2pt
		\resizebox{200pt}{120pt}{\includegraphics{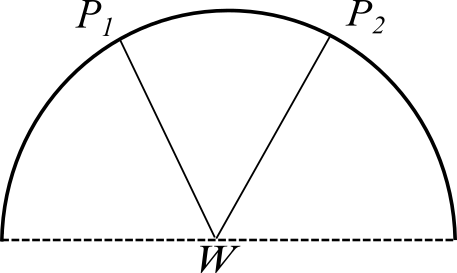}}
	\end{center}
	\caption{ The contact Witten diagram.}
\end{figure}%
For Dirichlet, we consider the vertex $\partial_{\perp}\Phi_1\partial_{\perp}\Phi_2$. There is a simplification here similar to the Dirichlet boundary exchange diagram. Following the notations of subsection \ref{bdysectn} , the contact diagram can be written as follows, 
\be\label{WContgenDir}
W_{\text{C}}^-=\int_{AdS_{d}}dW \  \partial_{\perp}^{W}G_{B\partial}^{-}(P_1,W|\D_1)\partial_{\perp}^{W}G_{B\partial}^{-}(P_2,W|\D_2)\,.
\ee
Due to \eqref{Dirichletprop} and the derivatives on the propagators, the above integral becomes equivalent to the Neumann case \eqref{WContgen} with effective external dimensions shifted as $\D_1\to \D_1+1$ and $\D_2\to \D_2+1$\,.

For Neumann boundary conditions, the boundary-bulk propagators are just the ordinary $AdS_{d+1}$ propagators (times $2$) . So the above integral simplifies to the interface case diagram computed in \cite{RastelliMellin}\,. For both boundary conditions, the contact diagram can be written in a compact way as follows,
\be\label{ContW}
\mathcal{W}_\text{C}^\nu(z)= \int_{-i \infty}^{i \infty}d\tau  \bigg(\frac{1-z}{z}\bigg)^{-\tau}  \frac{\G(\D_1-\tau+\frac{1-\nu}{2})\G(\D_2-\tau+\frac{1-\nu}{2})\G(\tau)}{\G\big(\frac{\D_1+\D_2-\nu}{2}-\tau+1\big)}\,.
\ee
As $z\to 0$ this has the decomposition,
\begin{align}
\mathcal{W}_{\text{C}}^\nu(z)&= z^{\frac{\D_1+\D_2}{2}} \sum_{n=0}^{\infty}\big[d_n^\nu(\D_1,\D_2) \hG_{\Delta_1+2n+\frac{1-\nu}{2}}(z|\D_1,\D_2)+d_n^\nu(\D_2,\D_1) \hG_{\Delta_2+2n+\frac{1-\nu}{2}}(z|\D_1,\D_2)\big] \nonumber \\ 
&\stackrel{\D_1= \D_2=\D_\phi}{=} z^{\D_\phi}\sum_{n=0}^{\infty}\bigg[d_n^{\nu} \hG_{\Df+2n+\frac{1-\nu}{2}}(z|\Df)+ e_n^{\nu} \partial_{\hat \Delta}\hG_{\Df+2n+\frac{1-\nu}{2}}(z|\Df)\bigg]\,,
\end{align}
where as before, when we approach $\Delta_1\to \Delta_\phi+\frac{\D_{12}}2$ and $\Delta_2\to\Delta_\phi-\frac{\D_{12}}2$, we have
\bea
d_n^{\nu}(\Delta_1,\Delta_2)\overset{\Delta_{12}\to 0}=\frac{ e_n^{\nu}}{\Delta_{12}}+\frac 12 d_n^{\nu}+O(\Delta_{12}),
\eea\\
As $z\to 1$ we have the bulk channel decompostion,
\bea
\mathcal W^{\nu}_{C}(z)&=& - z^{\frac{\Delta_1+\Delta_2}2}\sum_n f_n^\nu(\Delta_1,\Delta_2)G_{\Delta_1+\Delta_2+2n}^{\Delta_{12}}(z|\D_1,\D_2)
\nonumber \\
&\overset{\Delta_{12}=0}=& -z^{\Df}\sum_n f_n^\nu G_{2\Df+2n}(z|\Df)\,.
\eea
The expansion coefficients for the contact diagram are given in Appendix \ref{contdetails}\,.

\section{Polyakov and functional bootstrap}
\label{sec:polyakov}
\subsection{Polyakov blocks}
In section \ref{sec:functionals} we constructed two interesting bases of linear functionals that act on the BCFT crossing equation. This construction was partly motivated by considering the general structure of Witten diagrams in a bulk dual to the generalized free field BCFT. In the previous section we have explicitly computed these diagrams. We now close the circle and show in detail how these are related. 

The main claim of this section is that there exist Polyakov bulk and boundary blocks $\mathcal P_{\Delta}(z|\Df)$ and $\widehat{\mathcal P}_{\hD}(z|\Df)$, for both Neumann and Dirichlet boundary conditions, which satisfy:
\begin{subequations}
\label{eq:polyakov}
\bea
\mathcal P_\Delta(z|\Df)&=&G_{\Delta}(z|\Df)-\sum_{n=0}^{\infty} \theta_n(\Delta|\Df)G_{\Delta_n}(z|\Df)\nonumber \\
&=&\sum_{n=0}^{\infty}\left[ \alpha_n(\Delta|\Df) \hG_{\hD_n}(z|\Df)+\beta_n(\Delta|\Df) \partial_{\hD}\hG_{\hD_n}(z|\Df)\right]\\
\widehat{\mathcal P}_\hD(z|\Df)&=&\hG_{\hD}(z|\Df)-
\sum_{n=0}^{\infty}\left[ \alpha_n(\hD|\Df) \hG_{\hD_n}(z|\Df)+\beta_n(\hD|\Df) \partial_{\hD}\hG_{\hD_n}(z|\Df)\right]\nonumber \\
&=&\sum_{n=0}^{\infty} \theta_n(\hD)\,G_{\Delta_n}(z).
\eea
\end{subequations}
Concretely, we will show that the Polyakov blocks may be obtained from the Witten diagrams computed in the previous section. Indeed, note that the crossing relations satisfied by the Polyakov blocks look remarkably similar to those arising from Witten exchange diagrams, cf. \reef{eq:bulkdecomp}, \reef{eq:bdydecomp}, but with the notable difference that here the coefficients in the expansion are functional actions. We will show that it is possible to combine exchange diagrams with contact terms in a way that in those equations we effectively set:
\begin{subequations}
\label{eq:matching}
\bea
a_n(\Delta)&\to \alpha_n(\Delta),\qquad \hat a_n(\hD)\to \alpha_n(\hat \Delta)\\
b_n(\Delta)&\to \beta_n(\Delta),\qquad \hat b_n(\hD)\to \beta_n(\hat \Delta)\\
c_n(\Delta)&\to \theta_n(\Delta),\qquad \hat c_n(\hD)\to \theta_n(\hat \Delta)
\eea
\end{subequations}
In other words, the functional bases allow us to bootstrap Witten exchange diagrams.

Before we do this, let us briefly comment on why these Polyakov blocks are interesting objects. The role in life of the Polyakov blocks is that they provide us with a way of writing a generic two-point function in an explicitly crossing symmetric way. Concretely, for a given two point function $\mathcal G(z)$, we have:
\bea
\mathcal G(z)&=&\sum_{\hD} \mu_{\hD}^2 \hG_{\hD}(z)|\Df)=\sum_{\Delta} (a\lambda)_{\Delta}\,G_\Delta(z|\Df)\nonumber\\
 &=&\sum_{\hD} \mu_{\hD}^2 \widehat{\mathcal P}_{\hD}(z|\Df)+\sum_{\Delta} (a\lambda)_{\Delta}\,\mathcal P_\Delta(z)\label{eq:polyakovbs}
\eea
This remarkable statement is true only if the contributions from the unphysical bulk double trace operators and boundary derivative operators appearing in the conformal block decompositions of the Polyakov blocks drop out from the total sum. If we plug in the expression for the Polyakov blocks into the above equation and {\em commute the two series}, one finds that the unphysical states drop out precisely if the functional bootstrap equations \reef{eq:funcbootstrap} hold, which we repeat here for convenience:
\begin{subequations}
\begin{align}
\sum_{\hD} \mu_{\hD}^2 \alpha_n(\hD)&=\sum_{\Delta} (a\lambda)_{\Delta} \alpha_n(\Delta)\\
\sum_{\hD} \mu_{\hD}^2 \beta_n(\hD)&=\sum_{\Delta} (a\lambda)_{\Delta} \beta_n(\Delta)\\
\sum_{\hD} \mu_{\hD}^2 \theta_n(\hD)&=\sum_{\Delta} (a\lambda)_{\Delta} \theta_n(\Delta)
\end{align}
\end{subequations}
This establishes the link between the so-called Polyakov bootstrap and our functional bootstrap approach.

Of course, we have not shown that it is actually possible to commute the series. This would amount to showing that the functional bases we have constructed is in fact complete in some sense, i.e. that the functional bootstrap equations are completely equivalent to the original crossing equation. We will not attempt to prove this here. However, given the fact that these equations allow us to bootstrap arbitrary deformations of the generalized free solution, it seems likely that this is correct.\footnote{A similar problem arose in \cite{Mazac:2018ycv}. There the proof of completeness crucially relied on establishing upper bounds on the OPE density, so we expect the same must be true here. However, establishing such bounds seems more difficult here due to lack of positivity of the bulk channel expansion.}

\subsection{Dirichlet case}
Let us then show how Polyakov blocks and Witten diagrams are related, beginning with the Dirichlet case. In this case, we expect the situation to be simpler than for Neumann, since there were no ambiguities in the definition of our basis of functionals. As we've discussed before, there is a simple reason for this. 

Ambiguities in the basis correspond to the existence of solutions to crossing which preserve the form of the conformal block decomposition and which decay sufficiently fast at large $z$. This is crucial, since we have constructed our functional basis by assuming the functionals should satisfy the swapping property, but this is dependent on the large $z$ behaviour of a correlator. It can be checked by direct computation that the contact interaction with smallest number of derivatives and Dirichlet boundary conditions behaves like $\sim \sqrt{z}$ for large $z$. Since this is worse behaviour than for a generic two point function, which is only constant in the same limit, this explains the absence of ambiguities in this case, since such contact terms cannot be bootstrapped with our basis.\footnote{They can however be bootstrapped by demanding stronger falloff conditions on the functional basis.} 

On the other hand, in our computation of the  Witten diagrams we made specific choices of vertices on the brane. For Dirichlet boundary conditions it is straightforward to check that the boundary exchange Witten diagram falls off as $1/\sqrt{z}$ (as can be determined e.g. from its conformal block decomposition). The bulk exchange diagram also has the same fall-off behaviour. This can be checked by taking special values of the exchange dimension e.g. $\D=2\D_\phi-2$ .\footnote{In general for $\D=2\D_\phi-2m$ with integer $m$ the expression for $\mathcal{W}_\D^\nu$ simplifies (see \cite{RastelliMellin}).} Then we can infer that 1) we cannot shift the Witten exchange correlators by a contact interaction without changing the large $z$ behaviour, and 2) that there should be  no issue in acting directly with our functional basis on these correlators. By using the orthonormality properties of the functionals, it is natural to conclude:
\bea
z^{-\Df}\mathcal W^{-}_\Delta(z)&=&\mathcal P^-_{\Delta}(z)\\
z^{-\Df}\widehat{\mathcal W}^-_{\hD}(z)&=&\widehat{\mathcal P}^-_{\hD}(z).
\eea
That is, the Dirichlet functionals directly compute the conformal block decompositions of Dirichlet Witten exchange diagrams. To check this we should test relations \reef{eq:matching}. We have evaluated the functional actions for several $n$ using \reef{eq:funcactionbdy}, \reef{eq:funcactionbulk} and/or the original definition \reef{eq:funcactiondef}, as well as the kernels constructed in section \ref{sec:functionals}. To the extent we were able to determine these, the results are presented in appendix \ref{app:funcactions}. We then compared them with the conformal block decompositions of the Witten diagrams given in appendix \ref{app:witten1}. In some cases it is possible to do this analytically (in particular for the $\alpha,\beta$ funcitonals), in others it is more convenient to do numerical checks. In all cases we found perfect agreement.

\subsection{Neumann case}

The Neumann case is more interesting. Here the simplest contact interaction falls off as $z^{-\frac 12}$ near infinity, while the boundary exchange Witten  diagram computed in the section \ref{sec:witten} falls off like $z^{-\frac 32}$.\footnote{Recall the boundary channel decomposition in the Neumann case is related to the Dirichlet one by shifting everything by $\Delta_\phi\to \Df-1$ and dividing by $z$.} This is interesting for two reasons. Firstly, this means that our basis of functionals can act on any linear combination of the exchange diagram and the contact interaction, since it is suitable for correlators which grow as $z^{1/2-\eta}$ for any $\eta>0$. Secondly, recall that there is a missing Neumann functional precisely because we demanded that the falloff at infinity was sufficiently fast. This prefunctional was such that the decay at infinity was only $1/\sqrt{z}$. But we see that this is sufficient to act on the boundary exchange Witten  diagram, since overall we get a $1/z^2$ decay. The missing functional is simply%
\bea
\tilde \beta_0^+:\qquad h_+(z)=z\, h_{\beta_0^-}(z|\Df-1)
\eea
We put a tilde to emphasize these functionals have poorer asymptotic behaviour.
We can combine it with the remaining basis functionals constructed in section \ref{sec:functionals} to get a complete set of {\em fully} orthonormal prefunctionals, all of which decay at infinity as $z^{-\frac 12}$.  From the orthonormality relations \reef{eq:normalNeum} and \reef{eq:normalNeum2} the correct combination is
\begin{subequations}
\label{eq:tildefuncs}	
\bea
\tilde \alpha_n^+ &=& \alpha_n^++d_n\tilde \beta_0^+\nonumber\\
\tilde \beta_n^+&=& \beta_n^++e_n\tilde \beta_0^+,\nonumber\\
\tilde \theta_n^+&=& \theta_n^++f_n\tilde \beta_0^+.
\eea
\end{subequations}

The claim then is that these prefunctionals directly compute the conformal block decomposition of the Witten diagram, for example
\ba
z^{-\D_\phi}\widehat{\mathcal W}^+_{\hD}(z)&=\hG_{\hD}(z)-\sum_{n=0}^{\infty} \left[\tilde \alpha_n^+(\hD) \hG_{\Df+2n}(z)+\tilde \beta_n^+(\hD) \partial_{\hD}\hG_{\Df+2n}(z)\right]\,.
\ea
The $\tilde \alpha^+, \tilde \beta^+$ functional actions are determined in appendix \ref{app:funcactions}, and they can be indeed matched with the conformal block decompositions of the Neumann boundary exchange diagram. It works analogously for the bulk channel decomposition.

The bulk exchange diagram  \eqref{Wbulk} however falls off as $1/\sqrt{z}$, so the above prefunctionals cannot be used directly. One can however modify the  diagram by adding a contact diagram to \eqref{Wbulk}, since the resulting expression will still have the correct block decomposition as in \eqref{blockdeco1} and \eqref{blockdeco2}. Since the contact diagram also has the $1/\sqrt{z}$ fall-off, adding it with the right factor can cancel the leading large $z$ behavior, so that the modified diagram falls off as $z^{-\frac{3}{2}}$\,. We can write the modified bulk exchange diagram as ,
\be
\widetilde{\mathcal{W}}^+_{\Delta}(z)=\mathcal{W}^+_{\D}(z)-\frac{w_1}{w_2}\mathcal{W}^+_C(z)\,,
\ee
where the functions $w_1$ and $w_2$ are defined by the respective fall-off behaviours $z^{-\D_\phi}\mathcal{W}_\D^+\stackrel{z\gg 1}{\to} w_1z^{-\frac{1}{2}}$  and  $z^{-\D_\phi}\mathcal{W}_C^+\stackrel{z\gg 1}{\to} w_2z^{-\frac{1}{2}}$\,. The above object now allows the action of the prefunctionals \eqref{eq:tildefuncs}, and hence has a block decomposition,
\begin{align}
z^{-\D_\phi}\widetilde{\mathcal W}^+_{\D}(z)&=\sum_{n=0}^{\infty} \left[\tilde \alpha_n^+(\D) \hG_{\Df+2n}(z)+\tilde \beta_n^+(\D) \partial_{\hD}\hG_{\Df+2n}(z)\right]\nonumber\\
&=G_{\Delta}(z)-\sum_{n=0}^{\infty} \tilde \theta_n^+ (\Delta)G_{2\Df+2n}(z)\,.
\end{align}

We have not obtained the Polyakov block yet, since we want the coefficients to be the good functional basis with stronger falloff near $z=\infty$. Hence we expect that we need to take a suitable linear combination of the Witten exchange and the contact diagram to recover it. This combination is easily found. Since our basis does not contain $\beta_0^+$, we should have 
\bea
z^{\Df}\widehat{\mathcal P}_{\hD}^+(z)&=& \widehat{\mathcal W}_{\hD}^+(z) -\tilde \beta_0(\hD) \mathcal W_C^+(z) \\
z^{\Df} \mathcal P_{\Delta}^+(z)&=& \widetilde{\mathcal W}_{\D}^+(z) -\tilde \beta_0(\Delta) \mathcal W_C^+(z)\label{PDelta}\,.
\eea
It can be checked that for both the above objects the coefficient of $\partial_{\hD}\hG_{\Df}(z)$ cancels in the boundary conformal block expansions. This is compatible with \reef{eq:tildefuncs} if and only if $d_n, e_n, f_n$ compute the contact diagram, which we have already shown in the previous section. Note that we can alternatively define $\mathcal{P}_\D^+$ from $\mathcal W_{\D}^+$  by adding $\mathcal W_{C}^+$ and requiring the coefficient of $\partial_{\hD}\hG_{\Df}(z)$ to vanish from the boundary channel expansion (which is natural to expect given the decomposition of $\widehat{\mathcal{P}}^+_{\hat \D}$). This way one arrives at the same expression as above. It is straightforward to check that the condition \eqref{eq:matching} holds.

To conclude, as pointed out in section \ref{sec:functionals} our choice of basis of functionals is ambiguous in the Neumann case, since we cannot have fully orthonormal functionals. This ambiguity is reflected in the fact that unlike for Dirichlet boundary conditions, there is no longer a canonical Polyakov block because of the existence of the contact interaction. In particular, if $\omega$ is a good functional satisfying the correct fall-off condition, we are always free to shift:
\bea
\widehat{\mathcal P}_{\hD}^+&\to &\widehat{\mathcal P}_{\hD}^+ +\omega(\hD)z^{-\Df} \mathcal W_C(z). \\
\mathcal P_{\Delta}^+&\to & \mathcal P_{\Delta}^+ +\omega(\Delta)z^{-\Df} \mathcal W_C(z). 
\eea

\section{Application - Wilson-Fisher at $O(\epsilon^2)$}
\label{sec:WF}

The functional bootstrap equations \reef{eq:funcbootstrap} reformulate the crossing equation in a form which is ideally suited for perturbative expansions. As a particular application, we will reconstruct the BCFT data of the Wilson-Fisher fixed point, i.e. $\lambda \phi^4$ theory in $d=4-\epsilon$ dimensions tuned to the critical point) on a half-space with Neumann or Dirichlet boundary conditions, by bootstrapping the two point function of the fundamental field $\phi$. These boundary conditions preserve conformality on the defect and are usually called in the literature the {\em special} and {\em ordinary} transitions respectively. The usual CFT has been perturbatively studied extensively in literature \cite{Kleinert:1994td,Gopakumar2017,Alday:2016njk,Alday:2017zzv}, and one can consider the dimensions in bulk spectrum as input while doing BCFT bootstrap.

We will treat the Wilson-Fisher fixed point as a continuous deformation of the (generalized) free solution, i.e. one which preserves the structure of the spectrum of states in the two point function. This is certainly incorrect non-perturbatively since this spectrum is very different in both cases: for instance we expect that the number of operators appearing in the bulk channel increases exponentially fast for sufficiently large scaling dimension. In perturbation theory however, almost all such operators have coinciding scaling dimensions, and so we expect to get correct results for the $\langle \phi \phi\rangle$ correlator,  up to the order where degeneracies among operators should get lifted. For the Wilson-Fisher BCFTs this turns out to first happen at $O(\epsilon^3)$. More precisely at this order one will require extra information to disentangle the anomalous dimensions of degenerate operators but one should still recover the full correlator.\footnote{Beyond this order, the functional bootstrap equations do not fail, but rather they give rise to some other solution to crossing, namely one which has the sparsest possible spectrum of operators.The functional bootstrap equations give solutions to crossing, not theories, so it is quite likely such a solution is actually unphysical.} Here we will only consider perturbation theory up to $O(\epsilon^2)$.

The functional bootstrap equations take the form:
\bea
\sum_{n=0}^{\infty} \mu^2_{\hD_n}\, \omega_m(\hD_n|\Df,\epsilon)=\omega_m(0|\Df,\epsilon)+\sum_{n=0}^{\infty} (a\lambda)_{\Delta_n}\, \omega_m(\Delta_n|\Df,\epsilon)
\eea
where $\omega=\alpha,\beta,\theta$ and we've explicited the dependence of the functional actions on both $\Df$ and the spacetime dimension via $\epsilon$. We will expand parameters as follows:
\begin{equation}
\begin{aligned}
\label{eq:exp}
\hD_n&=\hD_n^{(0)}+\epsilon \hD_n^{(1)}+\epsilon^2 \hD_{n}^{(2)},&\qquad
\Delta_n&=\Delta_n^{(0)}+\epsilon \Delta_n^{(1)}+\epsilon^2 \Delta_{n}^{(2)},\\
\mu_n^2&=\mm_n^{(0)}+\epsilon \mm_n^{(1)}+\epsilon^2 \mm_n^{(2)},&\qquad
\al_n&=\te_n^{(0)}+\epsilon \te_n^{(1)}+\epsilon^2 \te_n^{(2)},\\
\Df&=\Delta_\phi^{(0)}+\epsilon \Delta_\phi^{(1)}+\epsilon^2 \Delta_\phi^{(2)}.
\end{aligned}
\end{equation}
with
\bea
\hD_n^{(0)}=\frac{1-\nu}2+\Delta_\phi^{(0)}+2n,\quad \Delta_n^{(0)}=2\Delta_\phi^{(0)}+2n,\quad \Delta_\phi^{(0)}=1.
\eea
It is also convenient to define
\bea
\hat \gamma_n^{(i)}=\hD_n^{(i)}-\Delta_\phi^{(i)},\quad \gamma_n^{(i)}=\D_n^{(i)}-2\Delta_\phi^{(i)},\qquad i=1,2,\ldots.
\eea
Note these are anomalous dimensions with respect to the {\em generalized} free field solution.

Upon inserting these expressions into the functional equations and expanding in $\epsilon$ we will obtain the functional actions and its derivatives with respect to $\Delta, \hD,\Df$ and $\epsilon$ evaluated at $\Df=1$ and $\epsilon=0$. We will omit this dependence below. 

\subsubsection*{$\bullet\,\,\, O(\epsilon^0)$}

At this order we merely recover the free solution. In particular we get
\begin{align}
\mm_n^{(0)}=\alpha_n^-(0)&=\delta_{n,0},& \te_n^{(0)}=-\theta_n^-(0)&=-\delta_{n0},& &\text{(Dirichlet)}\\
\mm_n^{(0)}=\alpha_n^+(0)&=2\delta_{n,0} ,& \te_n^{(0)}=-\theta_n^+(0)&=\delta_{n0},& &\text{(Neumann)}
\end{align}
These equations tell us that the free solution to crossing contains one block in both boundary and bulk channels, on top of the bulk identity. The fact that $\mm_n^{(0)},\t_n^{(0)}$ vanish for $n\geq 1$ will lead to simplifications below.

\subsubsection*{$\bullet\,\,\,O(\epsilon)$}
At this order we get:
\bea
&&\sum_{n=0}^{\infty}\left\{ \mm_n^{(1)} \omega_m(\hD_n^{(0)})+\mm_n^{(0)} \hat \gamma_n^{(1)} \partial_{\hD} \omega_m(\hD_n^{(0)}) \right\}\nonumber\\
&=&D_{\epsilon} \omega_m(0)+
\sum_{n=0}^{\infty}\left\{ \te_n^{(1)} \omega_m(\D_n^{(0)})+\te_n^{(0)} \gamma_n^{(1)} \partial_{\D} \omega_m(\D_n^{(0)}) \right\}.
\eea
In order to derive this expression we have used the orthonormality relations satisfied by $\alpha_m, \beta_m,\theta_m$ to simplify results, for instance, $\partial_\Df \omega_m(\D_n^{(0)})=-2 \partial_\D \omega_m(\D_n^{(0)})$. Using the results from the previous order we can get:
\begin{subequations}
\bea
\mm_m^{(1)}-D_\epsilon\alpha_m(0)&=&\te_0^{(0)}\gamma_0^{(1)}\partial_{\D}\alpha_m(\D_0^{(0)})-\mm_0^{(0)} \hat \gamma_0^{(1)} \partial_{\hD} \alpha_m(\hD_0^{(0)}) \\
\te_m^{(1)}+D_\epsilon\theta_m(0)&=&-\te_0^{(0)}\gamma_0^{(1)}\partial_{\D}\theta_m(\D_0^{(0)})+\mm_0^{(0)} \hat \gamma_0^{(1)} \partial_{\hD} \theta_m(\hD_0^{(0)}) \\
0&=&\te_0^{(0)} \gamma_0^{(1)} \partial_\D \beta_m(\D_0^{(0)})-\mm_0^{(0)} \hat \gamma_0^{(1)} \partial_{\hD} \beta_m(\hD_0^{(0)})
\eea
\end{subequations}
where we have defined:
\bea
D_{\epsilon}=\D_\phi^{(1)}\partial_{\D_\phi} +\partial_\epsilon.
\eea
On the other hand by explicit computation we find the relations:
\bea
\partial_\Delta \alpha_m(\Delta_0^{(0)})&=&-\partial_\hD\alpha_m(\hD_0^{(0)}), \qquad m\geq 1\\
\partial_\Delta \theta_m(\Delta_0^{(0)})&=&-\partial_\hD\theta_m(\hD_0^{(0)}), \qquad m\geq 2\\
\partial_\Delta \beta_m(\Delta_0^{(0)})&=&-\partial_\hD\beta_m(\hD_0^{(0)}), \qquad m\geq 1.
\eea
These expressions are valid for both Dirichlet and Neumann functionals. In the former case, the RHS of these equations is automatically zero. Recall also that all derivatives are evaluated for $\Df=1, \epsilon=0$. We find therefore
\ba
a_m^{(1)}&=0,& \qquad \text{for}\ m&\geq 1,\\
t_m^{(1)}&=0,& \qquad \text{for}\ m&\geq 2, 
\ea
for either boundary condition, where we have used that for the Wilson-Fisher fixed point we have $\D_\phi^{(1)}=-\frac 12$.  As for the remaining quantities we get %
\ba
a_0^{(1)}&=\frac 12 \gamma_0^{(1)}-\frac 12,&  t_0^{(1)}&=\frac 12\,\gamma_0^{(1)},& t_1^{(1)}&=\frac 14 \gamma_0^{(1)},& \hat \gamma_0^{(1)}&=-\frac 12 \gamma_0^{(1)},& &\text{(Dirichlet)}\\
a_0^{(1)}&=0,&  t_0^{(1)}&=\frac 12\,\gamma_0^{(1)},& t_1^{(1)}&=\frac 14 \gamma_0^{(1)},& \hat \gamma_0^{(1)}&=-\frac 12 \gamma_0^{(1)},&  &\text{(Neumann)}
\ea
This is in perfect agreement with the results of \cite{McAvity:1995zd,Liendo2013}.

\subsubsection*{$\bullet \,\,\,O(\epsilon^2)$}
At this order the equations become more involved.  We obtain:
\begin{equation}
\begin{split}
&\sum_{n=0}^{\infty}\bigg\{ \mm_n^{(2)} \omega_m(\hD_n^{(0)})+\left(\mm_n^{(0)} \hat \gamma_n^{(2)}+\mm_n^{(1)} \hat \gamma_n^{(1)}\right) \partial_{\hD} \omega_m(\hD_n^{(0)})+\\
&\qquad  +\frac 12 \mm_n^{(0)}\left[(\hD_n^{(1)})^2-(\Delta_\phi^{(1)})^2\right] \partial^2_{\hD} \omega_m(\hD_n^{(0)})+\mm_n^{(0)} \hat \gamma_n^{(1)} D_{\epsilon} \partial_{\hD} \omega_m(\hD_n^{(0)}) \bigg\}\\
=&\left(\frac 12 D^2_{\epsilon}+\D_{\phi}^{(2)} \partial_{\D_{\phi}}\right) \omega_m(0)+\sum_{n=0}^{\infty}\bigg\{ \te_n^{(2)} \omega_m(\D_n^{(0)})+
\left(\te_n^{(0)} \gamma_n^{(2)}+\te_n^{(1)} \gamma_n^{(1)}\right)\partial_{\D} \omega_m(\D_n^{(0)})\\
&\qquad \qquad +\frac 12 \te_n^{(0)} \left[(\D_n^{(1)})^2-4 (\D_{\phi}^{(1)})^2\right] \partial^2_\D \omega_m(\D_n^{(0)})+\te_n^{(0)} \gamma_n^{(1)}\, D_{\epsilon} \partial_\D \omega_m(\D_n^{(0)})\bigg\}
\end{split}
\end{equation}
where again we used the orthonormality properties of $\omega_m$ to simplify things.
Specializing to the $\alpha,\beta,\theta$ functionals and doing some simplifications we get
\begin{equation}
\begin{split}
&\mm_m^{(2)} +\frac 12 \mm_0^{(0)}(\hat\gamma^{(1)}_0)^2 \partial^2_{\hD} \alpha_m(\hD_0^{(0)})-
\\
&\qquad -\left(\mm_0^{(0)} \hat \gamma_0^{(2)}+\mm_0^{(1)} \hat \gamma_0^{(1)}\right) d_m  -\mm_0^{(0)} \hat \gamma_0^{(1)} D_{\epsilon} d_m \\
=&\D_{\phi}^{(2)} \partial_{\D_{\phi}}\alpha_m(0)+
\left(\te_0^{(0)} \gamma_0^{(2)}+\te_0^{(1)} \gamma_0^{(1)}\right)\partial_{\D} \alpha_m(\D_0^{(0)})+\te_1^{(1)} \gamma_1^{(1)}\partial_{\D} \alpha_m(\D_1^{(0)})
\\
&\qquad \qquad +\frac 12 \te_0^{(0)} \left[(\D_0^{(1)})^2-4 (\D_{\phi}^{(1)})^2\right] \partial^2_\D \alpha_m(\D_0^{(0)})+\te_0^{(0)} \gamma_0^{(1)}\, D_{\epsilon} \partial_\D \alpha_m(\D_0^{(0)})
\end{split}
\end{equation}
\vspace{0.2cm}
\begin{equation}
\begin{split}
&\mm_m^{(0)} \hat \gamma_m^{(2)}+a_m^{(1)}\hat \gamma_m^{(1)}
+\frac 12 \mm_0^{(0)}(\hat\gamma^{(1)}_0)^2 \partial^2_{\hD} \beta_m(\hD_0^{(0)})-
\\
&\qquad -\left(\mm_0^{(0)} \hat \gamma_0^{(2)}+\mm_0^{(1)} \hat \gamma_0^{(1)}\right) e_m  -\mm_0^{(0)} \hat \gamma_0^{(1)} D_{\epsilon} e_m \\
=&
\left(\te_0^{(0)} \gamma_0^{(2)}+\te_0^{(1)} \gamma_0^{(1)}\right)\partial_{\D} \beta_m(\D_0^{(0)})+\te_1^{(1)} \gamma_1^{(1)}\partial_{\D} \beta_m(\D_1^{(0)})
\\
&\qquad \qquad +\frac 12 \te_0^{(0)} \left[(\D_0^{(1)})^2-4 (\D_{\phi}^{(1)})^2\right] \partial^2_\D \beta_m(\D_0^{(0)})+\te_0^{(0)} \gamma_0^{(1)}\, D_{\epsilon} \partial_\D \beta_m(\D_0^{(0)})
\end{split}
\end{equation}
\vspace{0.5cm}
\begin{equation}
\begin{split}
&\frac 12 \mm_0^{(0)}(\hat\gamma^{(1)}_0)^2 \partial^2_{\hD} \theta_m(\hD_0^{(0)})-
\\
&\qquad -\left(\mm_0^{(0)} \hat \gamma_0^{(2)}+\mm_0^{(1)} \hat \gamma_0^{(1)}\right) f_m  -\mm_0^{(0)} \hat \gamma_0^{(1)} D_{\epsilon} f_m \\
=&\D_{\phi}^{(2)}  \partial_{\D_{\phi}}\theta_m(0)+\te_m^{(2)}+
\left(\te_0^{(0)} \gamma_0^{(2)}+\te_0^{(1)} \gamma_0^{(1)}\right)\partial_{\D} \alpha_m(\D_0^{(0)})+\te_1^{(1)} \gamma_1^{(1)}\partial_{\D} \theta_m(\D_1^{(0)})
\\
&\qquad \qquad +\frac 12 \te_0^{(0)} \left[(\D_0^{(1)})^2-4 (\D_{\phi}^{(1)})^2\right] \partial^2_\D \theta_m(\D_0^{(0)})+\te_0^{(0)} \gamma_0^{(1)}\, D_{\epsilon} \partial_\D \theta_m(\D_0^{(0)})
\end{split}
\end{equation}
In each of these three equations, note that the second line is only non-zero for the Neumann boundary condition. Recall that the coefficients $d_m, e_m, f_m$ provide the conformal block decomposition of the simplest contact interaction with Neumann boundary conditions, cf. appendix \ref{contdetails}. In this case, the second equation is also only valid for $m\geq 1$, as $\beta^+_0=0$.

Determining the second order coefficients is now a matter of evaluating derivatives of functional actions at specific values. In some cases this can be done analytically, but for the bulk functional actions especially this is difficult. For instance from the second set of equations one finds for Dirichlet boundary conditions:
\bea
\hat \gamma_0^{(2)}=-\gamma_0^{(2)}+\frac 14 \gamma_{0}^{(1)}\left(\gamma_{0}^{(1)}-1\right) \qquad \text{(Dirichlet)}
\eea

For the purposes of this work, we were content to check analytically where possible, and more generally numerically, that the results following from these equations are in perfect agreement with those of \cite{Bissi2018}.

\subsubsection*{$\bullet\,\,\,O(\epsilon^3)$ and beyond}

It is easy to see that there are no obstacles to going to higher orders in $\epsilon$ expansion. In particular at $p$-th order the equations look schematically as
\begin{equation}
\begin{split}
\sum_{n=0}^{\infty}\bigg\{ \mm_n^{(p)} \omega_m(\hD_n^{(0)})&+\mm_n^{(0)}\hat \gamma_n^{(p)} \partial_\hD \omega_m(\hD_n^{(0)})+(\ldots)\bigg\}\\
&=\D_\phi^{(p)} \partial_{\Df}\omega_m(0)+\sum_{n=0}^{\infty}\bigg\{ \te_n^{(p)} \omega_m(\D_n^{(0)})+\te_n^{(0)} \gamma_n^{(p)} \partial_{\D} \omega_m(\D_n^{(0)})+(\ldots)\bigg\}
\end{split}
\end{equation}
where on both sides of the equation the dots involve only results that are lower order in $\epsilon$. In particular, hidden in the dots we have contributions from infinite sets of operators in both bulk and boundary channels starting from $p=3$. Since the functionals satisfy orthornormality properties, it follows that by choosing $\omega=\alpha,\beta,\theta$ we can isolate specific $a_m^{(p)}, \hat \gamma_m^{(p)}$ or $t_m^{(p)}$ in the infinite sums above, e.g.:
\bea
a_n^{(p)}=\D_\phi^{(p)} \partial_{\Df} \omega_m(0)+\sum_{n=0}^{+\infty}\te_n^{(0)} \gamma_n^{(p)} \partial_{\D} \alpha_m(\D_n^{(0)})+(\ldots)
\eea
The order $\epsilon^p$ data will be expressed only in terms of results from previous orders in $\epsilon$, as well as the infinite set of bulk scaling dimensions $\gamma_n^{(p)}$ and $\Df^{(p)}$ which must be given as inputs.

\section{Discussion and outlook}

In this work we have constructed two functional bases for the BCFT crossing equation, which are associated to generalized free field solutions with Neumann and Dirichlet boundary conditions. The functionals lead to a set of sum rules, the functional bootstrap equations, which constrain the BCFT data. We have explained the relation of these equations to a Polyakov-like approach to the BCFT bootstrap, and used them to rederive perturbation theory results for the Wilson-Fisher fixed point to $O(\epsilon^2)$. There are a number of open problems and possible applications and extensions of these results.

The functional sum rules derived in this paper form a set of necessary conditions on the BCFT data. It would be nice to see if they can be used to derive bounds on this data as was done for 1D CFTs in \cite{Mazac:2018ycv}. We expect this to be possible at least in the boundary channel where there is manifest positivity. Such bounds are crucial if we want to show that the equations are also sufficient for crossing. In particular, until we do so, our proof of the validity of the Polyakov bootstrap will remain incomplete.

The sum rules lend themselves nicely not only to perturbative but also numerical studies of the crossing equation. This can be done by truncating the sum over the spectrum and considering only a finite set of functional equations. One can think of this as applying Gliozzi's \cite{Gliozzi2013} (or extremal flows \cite{El-Showk2018}) method  to the BCFT crossing equation in a particularly nice basis. Since the same basis reproduces the correct $\epsilon$-expansion coefficients, at least to some order, it is reasonable to expect that repeating the analysis numerically should also give good results. Right now the chief technical obstacle both for perturbative and numerical analysis is better control over the functional actions.


We have discussed two discussed kinds of bases for the BCFT crossing equation and their associated functionals, in sections \ref{sec:funcans} and \ref{sec:functionals}. For both kinds, generic dimension bulk and boundary blocks were expanded in terms of double trace bulk blocks and boundary derivative blocks. In the first, ``easy'' basis, the boundary channel was comprised of blocks $\hG_{\hD_n}$ of dimension $\Delta_n=\Delta_{\hat \phi}+n$ for $n=0,1,\ldots$, whereas in the main bases of section \ref{sec:functionals} we had $\Delta_n=\Delta_{\hat \phi}+2n$. This halving was compensated by the appearance of derivatives of boundary blocks $\partial_{\hD} \hG_{\hD_n}$. It was thanks to the functionals dual to these latter objects (i.e. the $\beta_n$) that we were able to control deformations of the generalized free solution where the boundary operator dimensions were allowed to change.

In both kinds of bases however, the bulk channel is comprised of blocks with dimension $G_{\Delta_n}$ with $\Delta_n=2\Df+2n$, but not their derivatives. This implies that in this basis we do not have good control over deformations where the bulk operator dimensions change, which means in practice these must be inputs. It seems then that there should be a third kind of basis, where also in the bulk channel we halve the set of operators and introduce compensating $\partial_{\Delta} G_{\Delta_n}$. In general however, we expect that such a basis should be very non-trivial, since it would not be associated to generalized free fields. The reason why we should expect this basis to exist at all is because the associated functionals {\em can} be constructed numerically: indeed, the original numerical bootstrap for BCFTs implemented in \cite{Liendo2013} can be thought of as constructing exactly such functionals. It would be very interesting to see what can be said of such basis, as they might then allow us to access scaling dimensions of higher-d bulk CFTs.

In this paper we have constructed explicit expressions for Witten diagrams for BCFTs, subject to Neumann and Dirichlet boundary conditions.  It would be worthwhile to explore their structure more, to see if there are further simplifications  and interesting features in their block decomposition \cite{Gopakumar:2018xqi, Zhou:2018sfz}. 

Let us conclude by noting there are several possible generalizations of the results in this paper which should be obtainable with moderate efforts. Among these stand out bootstrapping two point functions of spinning fields \cite{Lauria:2018klo} and including various amounts of supersymmetry \cite{Liendo:2016ymz}. The latter is particularly interesting in light of applications to line defects in superconformal CFTs where many exact results are available. We look forward to exploring these and other developments in the future.

\vspace{1 cm}
\section*{Acknowledgments}
We would like to thank Adel Bilal, Parijat Dey, Kausik Ghosh, Rajesh Gopakumar, Dalimil Maz\'a\v{c}, Leonardo Rastelli, Slava Rychkov, Aninda Sinha, Emilio Trevisani and Xinan Zhou for useful discussions and/or comments on the draft. AK and MFP acknowledge the role of the Simons bootstrap collaboration workshops in Azores and Caltech where parts of this work were completed. AK would like to thank YITP, Stony Brook for hospitality during the course of this work. AK is supported by the Simons Foundation grant 488655 (Simons Collaboration on the Nonperturbative Bootstrap).
\pagebreak

\appendix

\section*{Appendix}

\section{Functional Actions}
\label{app:funcactions}
In this section we compute the functional actions starting from the expressions for the kernels determined in section \ref{sec:functionals}

\subsection{$\alpha$ and $\beta$ functionals}

The $\alpha$ and $\beta$ functionals can be obtained from the $\omega_{m,k}$ by Gram-Schmidt decomposition. Concretely, we have:
\bea
\beta_n^-&=&\frac{\omega_{2n,0}-\sum_{i=0}^{n-1} \partial_{\hD} \omega_{2n,0}(\hD_{i}^-) \beta_{i}^-}{\partial_{\hD} \omega_{2n,0}(\hD_n^-)}\\
\beta_n^+&=&\frac{\omega_{2n-1,0}-\sum_{i=1}^{n-1} \partial_{\hD} \omega_{2n-1,0}(\hD_{i}^+) \beta_{i}^+}{\partial_{\hD} \omega_{2n-1,0}(\hD_n^+)}\\
\alpha_n^-&=&\frac{\omega_{2n+2,2n+2}-\sum_{i=0}^{n+1} \partial_{\hD} \omega_{2n+2,2n+2}(\hD_{i}^-) \beta_{i}^--\sum_{i=1}^{n-1} \omega_{2n+2,2n+2}(\hD_{i}^-) \alpha_i^-}{\omega_{2n+2,2n+2}(\hD_n^-)}
\\
\alpha_n^+&=&\frac{\omega_{2n+1,2n+1}-\sum_{i=1}^{n+1} \partial_{\hD} \omega_{2n+1,2n+1}(\hD_{i}^+) \beta_{i}^+-\sum_{i=1}^{n-1} \omega_{2n+1,2n+1}(\hD_{i}^+) \alpha_i^-}{\omega_{2n+1,2n+1}(\hD_n^+)}
\eea
with $\hD_n^-=1+\Df+2n$ and $\hD_n^+=\Df+2n$.

\subsubsection{Action on boundary blocks}
One finds:
\bea
\omega_{m,k}(\hD)=\scalebox{1}{$-\frac{4^{-\Delta _{\phi }+\Delta -m-\frac 12} \Gamma \left(\Delta +\frac{3-d}2\right) \Gamma \left(k+\Delta _{\phi }+\frac{3-d}{2}\right) \Gamma \left(m+\Delta _{\phi
   }+1\right)}{(m-k)! \Gamma (\Delta )  \Gamma \left(\frac{m-\Delta +\Delta _{\phi }+3}2\right) \Gamma \left(\frac{m+\Delta
   +\Delta _{\phi }+4-d}2\right) \Gamma \left(\frac{2 k-m-\Delta +\Delta _{\phi }+1}2\right) \Gamma \left(\frac{2 k-m+\Delta
   +\Delta _{\phi }+2-d}2\right)}$}
\eea
Using this expression we find
\bea
\beta^+_n(\hD)=\tilde \beta^+_n(\hD)-e_n \tilde \beta^+_0(\hD)
\eea
where
\bea
\tilde \beta^+_n(\hD)=\scalebox{1}{$\frac{4^{ \hD -\hD _n+\frac 32} \Gamma \left(-\frac{d}{2}+\hD +\frac{3}{2}\right) \Gamma \left(\hD _n\right) \Gamma \left(\frac{-d+\hD
   _n+\hD _{0}+1}2\right)^2}{\Gamma (\hD ) \left(\hD -\hD _n\right) \left(-d+\hD +\hD _n+1\right) \Gamma \left(\frac{\hD
   _{0}-\hD }2\right)^2 \Gamma \left(\frac{-d+\hD +\hD _{0}+1}2\right)^2 \Gamma \left(-\frac{d}{2}+\hD
   _n+\frac{1}{2}\right) \Gamma \left(\frac{\hD _n-\hD _{0}+2}2\right)^2}$} \label{eq:betatilde}
	\eea
In this expression $\hD_n=\Df+2n$. Recall that the $\tilde \beta, \tilde \alpha$ are the fully orthonormal Neumann prefunctionals, i.e. functionals whose kernels fall off as $z^{-\frac 12}$ at infinity, and not as $z^{-\frac 32}$ as they should. The $e_n$ are found to be
\bea
e_n=\frac{4^{\Delta _{\phi }-\hD_n} \Gamma \left(\hD_n\right) \Gamma \left(\frac{-d+\hD_n+\Delta _{\phi }+1}2\right)^2}{\Gamma \left(\Delta
   _{\phi }\right) \Gamma \left(-\frac{d}{2}+\Delta _{\phi }+\frac{1}{2}\right) \Gamma \left(-\frac{d}{2}+\hD_n+\frac{1}{2}\right) \Gamma \left(\frac{\hD_n-\Delta _{\phi }+2}2\right)^2}.
\eea
The Dirichlet functionals are then obtained simply by
\bea
\beta_n^-(\hD|\Df)=\tilde \beta_n^+(\hD|\Df+1).
\eea
In particular this leads to exactly the same expression as the RHS of \reef{eq:betatilde} but now with $\hD_n=1+\Df+2n$.

As for the $\alpha_n$ functionals, after computation we observe that:
\bea
\alpha_n=\frac 12 \partial_n \beta_n.
\eea
This holds for tilded and untilded Neumann functional actions as well as Dirichlet. In particular we also get
\bea
d_n=\frac 12 \partial_n e_n.
\eea

\subsubsection{Acting on bulk blocks}
In this case we find
\bea
\omega_{m,k}(\Delta)=\scalebox{1}{$
 N_{m,k}(\Delta)\sin \left[\frac{\pi}2(\Delta-2\Df)\right] \, _4\tilde{F}_3
	\bigg[\scalebox{0.75}{$  \!\!\!\!\!\!\begin{matrix} &\frac{\Delta -2 \Delta _{\phi
   }+2}2, \frac{\D}2 , \frac{\D}2,k+\frac{1+\D}2 \\ &
 \frac{4 k-2 m+\Delta }{2}, m+\frac{\Delta }{2}+2,\frac{2-d}{2}+\Delta \end{matrix};1$} \bigg]    
$}
\eea
with 
\bea
N_{m,k}(\Delta)=
\frac{4^{k-m} \Gamma \left(\frac{2-d}{2}+\Delta \right) \Gamma \left(\frac{\Delta }{2}-\Delta _{\phi
   }+1\right) \Gamma \left(k+\frac{\Delta +1}{2}\right) \Gamma \left(m+\Delta _{\phi }+1\right)    
}{\pi ^{3/2}
   (m-k)!}.
\eea
After Gram-Schmidt decomposition we were able to find relatively simple expressions for the $\beta_n(\D)$:
\bea
\beta_n^-(\D)&=&\sin \left[\frac{\pi}2(\Delta-2\Df)\right]\sum_{k=0}^n b_{n,k}^-(\Delta) _4\tilde{F}_3
	\bigg[\scalebox{0.75}{$  \!\!\!\!\!\!\begin{matrix} &\frac{\Delta -2 \Delta _{\phi
   }+2}2, \frac{\D}2 , \frac{\D}2,\frac{1+\D}2 \\ &
 \frac{-4 k+\Delta }{2},\frac{4+4k+\Delta }{2},\frac{2-d}{2}+\Delta \end{matrix};1$} \bigg] \\
\beta_n^+(\D)&=&\sin \left[\frac{\pi}2(\Delta-2\Df)\right]\sum_{k=1}^n b_{n,k}^+(\Delta) _4\tilde{F}_3
	\bigg[\scalebox{0.75}{$  \!\!\!\!\!\!\begin{matrix} &\frac{\Delta -2 \Delta _{\phi
   }+2}2, \frac{\D}2 , \frac{\D}2,\frac{1+\D}2 \\ &
 \frac{2-4 k+\Delta }{2},\frac{2+4k+\Delta }{2},\frac{2-d}{2}+\Delta \end{matrix};1$} \bigg]
\eea
with
\bea
b_{n,k}^-(\Delta)&=&\frac{(1+2k)}{4^{2n}\pi^{\frac 32}}\frac{\left(\Df+\frac {1-d}2\right)_{n-k}}{\left(\Df+n+k+\frac {3-d}2\right)_{n-k}}\scalebox{1}{$\frac{\Gamma\left(\hD^-_n\right)\G\left(1-\Df+\frac{\D}2\right)\G\left(\frac{1+\Df}2\right)\G\left(\D+\frac{2-d}2\right)}{(n-k)!(n+k+1)!}$}\\
b_{n,k}^+(\Delta)&=&\frac{2k}{4^{2n-1}\pi^{\frac 32}}\frac{\left(\Df+\frac {1-d}2\right)_{n-k}}{\left(\Df+n+k+\frac {1-d}2\right)_{n-k}}\scalebox{1}{$\frac{\Gamma\left(\hD^+_n\right)\G\left(1-\Df+\frac{\D}2\right)\G\left(\frac{1+\Df}2\right)\G\left(\D+\frac{2-d}2\right)}{ (n-k)!(n+k)!}$}
\eea
To obtain the tilded functionals one just needs to know that
\bea
\tilde \beta_0=\omega_{-1,-1},
\eea
since $\tilde \beta_n=\beta_n+e_n \tilde \beta_0$.

As for the $\alpha_n(\Delta)$ functional actions, these two are given in terms of finite sums of $_4F_3(1)$ as above, but we were now not able to derive a closed form expression for their coefficients.

\subsection{Bulk functionals}
The orthonormal bulk functionals may be obtained from the shifted bulk functionals by Gram-Schmidt orthonormalization, which in this case takes the form
\bea
\theta_n^-&=&s\theta_n-\sum_{i=0}^{n-1} s\theta_n^-(\Delta_i) \theta_{i}^-\\
\tilde \theta_n^+&=&s\theta_n^+-\sum_{i=0}^{n-1} s\theta_n^+(\Delta_i) \tilde \theta_{i}^+
\eea
To recover the untilded functionals, i.e. those with the correct falloff in the Neumann case there is a slight modification:
\bea
\theta_n^+&=&s\theta_n^+-\sum_{i=0}^{n-1} s\theta_n^+(\Delta_i) \theta_{i}^+, \qquad n\geq 1\\
\theta_0^+&=&\tilde \theta_0^+-f_0 \tilde \beta_0,\qquad \tilde \theta_0^+=s\theta_0^+
\eea
where $f_0$ was previously determined to be
\bea
f_0=\frac{\sqrt{\pi} \Gamma(\Df)}{\Gamma\left(\frac 12+\Df\right)}
\eea
Note we must have also%
\bea
\theta^+_n=\tilde \theta^+_n-f_n \tilde \beta_0
\eea
where
\be
f_n=\frac{(-1)^{n}\sqrt{\pi} (2\D_\phi+1-d) \Gamma (n+\text{$\Delta_\phi $})^2 \left( \D_\phi-h+\frac{3}{2}\right)_{n-1}}{2 (n!) \Gamma(\D_\phi)\Gamma \left( n+\D_\phi+\frac{1}{2}\right) \left(n+2\text{$\Delta_\phi $}-h\right)_n}\,.
\ee
is determined in appendix \ref{contdetails}. Consistency requires
\bea
\sum_{i=0}^n s\theta_n^+(\Delta_i)f_i=0
\eea
and we have checked this is the case numerically to great accuracy.

As for the shifted functional actions we find
\bea
s\theta^{-}_n(\Delta)=\scalebox{1}{$\sin \left[\frac{\pi}2(\Delta-2\Df)\right]$} \mathcal N_n^-\sum_{p=0}^{+\infty}\sum_{i=0}^1 t_{n,p,i}^-(\Delta)
_3\tilde{F}_2
	\bigg[\scalebox{0.75}{$  \!\!\!\!\!\!\begin{matrix} &p-\frac 12,\frac{\D+2-d}2,i+\frac{2\Df+2n+2p-1-d}2 \\ &
 \frac{\Delta+1-d+p}2,i+\frac{2\Df+\Delta+2n+2p-1-d}2 \end{matrix};1$} \bigg]\\
s\theta^{+}_n(\Delta)=\scalebox{1}{$\sin \left[\frac{\pi}2(\Delta-2\Df)\right]$} \mathcal N_n^+\sum_{p=0}^{+\infty}\sum_{i=0}^1 t_{n,p,i}^+(\Delta)
_3\tilde{F}_2
	\bigg[\scalebox{0.75}{$  \!\!\!\!\!\!\begin{matrix} &p-\frac 32,\frac{\D+2-d}2,i+\frac{2\Df+2n+2p-3-d}2 \\ &
 \frac{\Delta-1-d+p}2,i+\frac{2\Df+\Delta+2n+2p-3-d}2 \end{matrix};1$} \bigg]
\eea
%
and
\bea
s\theta^{-}_n(\hD)=\scalebox{1}{$\sin^2 \left[\frac{\pi}2(\Delta-\hD_0^-)\right]$} \widehat {\mathcal N}_n^-\sum_{p=0}^{+\infty}\sum_{i=0}^1\widehat{ t}_{n,p,i}^-(\hD)
_3\tilde{F}_2
	\bigg[\scalebox{0.75}{$  \!\!\!\!\!\!\begin{matrix} & \D,\Df+\D+1-d,\frac{2\D+2-d}2 \\ &
 2\D+2-d,i+\Df+\D+n+p+\frac{1-d}2 \end{matrix};1$} \bigg]\\
s\theta^{+}_n(\hD)=\scalebox{1}{$\sin^2 \left[\frac{\pi}2(\Delta-\hD_0^+)\right]$} \widehat{\mathcal N}_n^+\sum_{p=0}^{+\infty}\sum_{i=0}^1 \widehat{t}_{n,p,i}^+(\hD)
_3\tilde{F}_2
	\bigg[\scalebox{0.75}{$  \!\!\!\!\!\!\begin{matrix} & \D,\Df+\D-d,\frac{2\D+2-d}2 \\ &
 2\D+2-d,i+\Df+\D+n+p-\frac{1+d}2 \end{matrix};1$} \bigg]
\eea
The coefficients appearing in these and the above expressions can be obtained from the authors upon request.

\section{Computation of bulk exchange Witten diagram}\label{bulkdetails}
In this appendix we give the details of the computation of the bulk Witten diagram \eqref{2intgrl}, which we show here for convenience,
\bea
W_\Delta^\nu(P_1,P_2)=\int_{AdS_{d+1}}dX\int_{AdS_{d}}dW \Big[ G_{B\partial}(P_1,X|\D_1)G_{B\partial}(P_2,X|\D_2)G_{BB}(X,W|\D)+\nonumber\\
+\nu\, G_{B\partial}(P_{1r},X|\D_1)G_{B\partial}(P_2,X|\D_2)G_{BB}(X,W|\D)\Big]\,.
\eea
The first part of \eqref{2intgrl} is the same bulk exchange integral evaluated in \cite{RastelliMellin}, for the interface boundary condition\,. The second integral is slightly different, so let us work it out in detail. The bulk-boundary propagator $G_{B\partial}(P,X)$ is given by \eqref{bulkbdyprop}. The bulk-bulk propagator in its spectral representation is given by,
				\be
				G_{BB}(X,W|\D)=\int_{-i\infty}^{i\infty}\frac{dc}{(c^2-(\D-h)^2)}\frac{\Gamma(h+c)\Gamma(h-c)}{2\pi^{2h}\Gamma(c)\G(-c)}\int_{\partial AdS_{d+1}}dP (-2P.X)^{h+c}(-2P.W)^{h-c} \,.
				\ee
				Carrying out the integral over $X$, which is over bulk $AdS_{d+1}$ gives us the following three-point function ,
				\begin{align}
				\la O_1 & (P_1)O_2(P_{2r})  O_{h+c}(P)  \ra = \nonumber \\&
\frac{\pi^h\G\left(\frac{\D_1+\D_2-h-c}{2}\right)\G\left(\frac{\D_1+\D_2-h+c}{2}\right)\G\left(\frac{\D_1-\D_2+h+c}{2}\right)\G\left(\frac{\D_2-\D_1+h+c}{2}\right)}{2\G(\D_1)\G(\D_2)\G(h+c)(-2P_1.P_{2r})^{\frac{\D_1+\D_2-h-c}{2}}(-2P_1.P)^{\frac{h+c+\D_{12}}{2}}(-2P_{2r}.P)^{\frac{h+c+\D_{21}}{2}}}\,.
				\end{align}
				Here we have
\begin{align}
-2P_1.P_{2r}&=({\vec x}_1-{\vec x}_2)^{2}+({ x_{1}^\perp}+{x_{2}^\perp})^2\nonumber \\
-2P_1.P&=({\vec x}_1-{\vec x})^{2}+({ x_{1}^\perp}-{x^\perp})^2\nonumber \\
-2P_{2r}.P&=({\vec x}_2-{\vec x})^{2}+({ x_{2}^\perp}+{x^\perp})^2\,.
\end{align} 
Carrying out the $W$ integral gives the one point function,
				\be
				\la O_{h-c}(P)\ra= \frac{\pi^{h-\frac{1}{2}}\Gamma\left(\frac{h-c}{2}\right)\G(\frac{-h-c+1}{2})}{2\G(h-c)({x^\perp}^ 2)^{\frac{h-c}{2}}}\,.
				\ee
				The integral over $P$ can be done with Schwinger parameters, that yields,
{\small
				\begin{align}
				\int_{-\infty}^{\infty} &   \frac{dx^{\perp}  d^{d-1}x}{(-2P_{2r}.P)^{\frac{h+c+\D_1-\D_2}{2}}(-2P_1.P)^{\frac{h+c+\D_2-\D_1}{2}}({x^\perp}^ 2)^{\frac{h-c}{2}}}\nonumber = \frac{\pi^h}{\G\Big(\frac{h+c+\D_{12}}{2}\Big)\G\Big(\frac{h+c+\D_{21}}{2}\Big)\G\Big(\frac{h-c}{2}\Big)} \\ & \times \int_0^{\infty}\frac{ds}{s}\frac{dt}{t}\frac{du}{u}\frac{s^{\frac{h-c}{2}}t^{\frac{h+c+\D_{12}}{2}}u^{\frac{h+c+\D_{21}}{2}}}{(t+u)^{h-\frac{1}{2}}(s+t+u)^{\frac{1}{2}}}   \exp\Big[-\frac{t u (-2P_1.P_{2r})}{t+u} - \frac{s (t x_1^\perp-u x_2^\perp)}{(t+u)(s+t+u)}\Big]\,.
				\end{align}
}
				Here $x_1^\perp,x_2^\perp>0$\,. To evaluate the integral, the usual steps are to insert the following,
				\be
				1=\int d\lambda \ \delta(\lambda-(s+t+u))\int d\rho \  \delta(\rho-(t+u))\,,
				\ee
				and do the $s$ integral. Then we rescale $\lambda=\rho\lambda$, $t=\rho t$ and $u=\rho u$ and carry out the $t$ integral\,. This gives,
				\begin{align}
				\int_1^{\infty} d\lambda \int_0^{1}du & \int_0^{\infty}d\rho  \  (\lambda-1)^{\frac{h-c}{2}-1}(1-u)^{\frac{h+c+\D_{12}}{2}-1}u^{\frac{h+c+\D_{21}}{2}-1} \rho^{\frac{h+c}{2}-1}  \nonumber \\ &\times \exp\Big[-\rho u (1-u) (-2P_1.P_{2r}) - \frac{\rho(\lambda-1) }{\lambda}((1-u) x_1^\perp-u x_2^\perp)^2\Big]\,.
				\end{align}
				Note that the limits of the integrals follow from the above rescalings. Here we have to do an inverse  Mellin transform, to rewrite one part inside the exponential as a power law. For this we use the following,
				\be
				e^{-x}=\int_{-i \infty}^{i\infty}\G(\tau)x^{-\tau}d\tau\,.
				\ee
				We choose the arguments of exponentials and power laws carefully such that the integrals over $\lambda$ and $u$  can be carried out. Rearranging the exponential slightly we can write,
				\begin{align}
				\exp & \Big[-\rho u (1-u) (-2P_1.P_{2r}) - \frac{\rho(\lambda-1) }{\lambda}((1-u) x_1^\perp+u x_2^\perp)^2 + 4\frac{\rho(\lambda-1) }{\lambda}(u (1-u)x_1^\perp x_2^\perp)\Big]\nonumber\\
				=&\exp\Big[-\rho u (1-u) (-2P_1.P_{2r})  + 4\frac{\rho(\lambda-1) }{\lambda}(u (1-u)x_1^\perp x_2^\perp)\Big]\nonumber\\ &\times \int_{-i\infty}^{i\infty}d\tau\G(\tau)\left( \frac{\rho(\lambda-1) }{\lambda}((1-u) x_1^\perp+u x_2^\perp)^2\right)^{-\tau}\,.
				\end{align}
				One can easily verify that the quantities inside exponential and in power of $\tau$ are of  definite signs, which allows us to do the $\rho, u$ and $\lambda$ integrals. This gives a hypergeometric ${}_2F_1$. The final answer, including the first term of \eqref{2intgrl}, is given by \eqref{Wbulk}\,.

\section{Witten diagram block decompositions}
\label{app:witten1}

In this appendix we are going to give the details of the coefficients of expansion of the various Witten diagrams, and how to compute them. Since the Witten diagram expressions in section \ref{sec:witten} are written as functions of $(1-z)/z$, it is convenient to use the variable $\xi$, 
\be
\xi=\frac{1-z}{z}\,.
\ee
In terms of this, the bulk and boundary blocks are respectively given by,
\begin{align}\label{oldnewblock}
G_{\D}(z|\D_1,\D_2)&=\xi^{\frac{\D-\D_1-\D_2}{2}}(1+\xi)^{\frac{\D_1+\D_2}{2}}{}_2F_1(\frac{\D+\D_{12}}{2},\frac{\D+\D_{21}}{2},\D-h+1,-\xi)\nonumber \\
\hG_{\hat \D}(z|\D_1,\D_2)&=\xi^{-\hat \D}(1+\xi)^{\frac{\D_1+\D_2}{2}}{}_2F_1\big(\hat \D,\hat \D-h+1,2\hat \D+2-d,-\frac{1}{\xi}\big)\,.
\end{align}
It is clear that the limit $z\to 0$, which corresponds to boundary channel, will be given by $\xi \to \infty$\,. In the bulk channel, where $z\to 1$, we have $\xi\to 0$\,.

\subsection{Boundary exchange diagram}\label{bdyExcdetails}

In the diagram \eqref{WbdyNm} with Neumann boundary conditions, we have the following set of poles in $\tau$,
\begin{enumerate}
\item {$\tau=h'\pm c+m_1$, for $m_1 \in \mathbb{Z}$ and $m_1\ge 0$}

\item{$\tau=-m_2$, for $m_2 \in \mathbb{Z}$ and $m_2 \ge 0$}
\end{enumerate}
and the following poles in $c$,
\begin{enumerate}
\item {$c=\pm(\hat\D-h')$}

\item{$c=\pm(\D_1-h'+2m_3)$ for $m_3 \in \mathbb{Z}$ and $m_3\ge 0$}

\item{$c=\pm(\D_2-h'+2m_4)$ for $m_4 \in \mathbb{Z}$ and $m_4\ge 0$}\label{3rdset}

\item{$c=\pm(h'- \tau+m_5)$, for $m_5 \in \mathbb{Z}$ and $m_5 \ge 0$}
\end{enumerate}

In the boundary channel decomposition \eqref{bdydeco2}, we have $\xi \to \infty$\,.  Here the various $\xi$ dependencies arise from the collisions of $\tau$ and $c$ poles. The $\tau$-contour  is closed on the right. To obtain the ``physical term", we take the residues at the poles $\tau=h'\pm c$. For the `$+$' (`$-$') sign we close the $c$-contour on the right (left) and then choose $c=\hat\D-h'$ \ ($c=h'-\hat\D$). The total gives us the first term in the large $\xi$ expansion of the boundary block $\widehat G_{\hat \D}((1+\xi)^{-1}|\D_1,\D_2)$. Since we normalize it to unit coefficient, this gives us $\mathcal{N}^{+}_{\text{bdy}}(\hat\D)$ that reads,
\be
\mathcal{N}^{+}_{\text{bdy}}(\hat\D)=\frac{4^{\Delta +1}\Gamma (\hat\Delta +\frac{3}{2}-h)}{ \Gamma (\hat\Delta ) \Gamma (\frac{\Delta _1-\hat\Delta }{2} ) \Gamma (\frac{\Delta _2-\hat\Delta}{2} ) \Gamma (\frac{\hat\Delta +\Delta _1+1}{2} -h) \Gamma (\frac{\hat\Delta +\Delta _2+1}{2} -h)}\,.
\ee
Let us now compute the coefficient of the boundary block $\widehat G_{\D_1+2n}$, which is related to the coefficients of $\xi^{-\D_1-2n}$ .  It is obtained by taking residues at $\tau=h'\pm c+m_1$. Once again the choice of the $c$-contour depends on the sign. Then we take residues at $ c=\D_1-h+2\big[n-\frac{m_1}{2}\big]$ with $m_1=1,2, \cdots n$ (or $ c=-\D_1+h'-2\big[n-\frac{m_1}{2}\big]$ as appropriate from the choice of contour). This gives all the contribution to $\xi^{-\D_1-2n}$, from which we subtract the contributions from the blocks $\widehat G_{\D_1+2k}$ with $k<n$. This gives us the coefficient   $a_n^+  (\D_1,\D_2,\hat\D)$ ,
\be\label{ancoeff}
{\scalebox{1.04}{ $\hat a_n^+  (\D_1,\D_2,\hat\D)=\mathcal{N}^{+}_{\text{bdy}}(\hat\D) \frac{(-1)^{n} \Gamma (2 n+\Delta _1) \Gamma (\frac{\Delta _2-2 n-\Delta _1}{2} ) \Gamma (n+\Delta _1+\frac{1}{2}-h) \Gamma (\frac{\Delta _1+\Delta _2+1}{2} -h+ n)}{ 4^{ \Delta _1+2 n} n!(\hat\Delta -\Delta _1-2 n) (\hat\Delta +\Delta _1+2 n+1-d) \Gamma (2 n+\Delta _1+\frac{1}{2}-h)}$}}\,.
\ee
The third set of poles in $c$ simply gives us $\hat a_n^+  (\D_2,\D_1,\hat\D)$ similarly as above. The fourth set has already been used up from the $\tau$ poles. For equal scalars the coefficients $\hat b_n^+ $ and $\hat a_n^+ $ can be obtained from \eqref{ancoeff} as shown in the main text.

In the bulk channel expansion of the diagram \eqref{WbdyNm}, we have $\xi \to 0$. So the $\tau$-contour is closed on left. The coefficient $\hat t^{+}_n$ of the bulk block $G_{\D_1+\D_2+2n}$ can be obtained by taking the residues at the poles $\tau=-m_2 $ of $\Gamma(\tau)$ and rearranging  the powers of $\xi$ as expected in the bulk blocks . The coefficient of $\xi^n$, coming from the residue at $\tau=-n$ is given by,
\be
\tilde {\hat t}_n^+  (\D_1,\D_2,\hat\D)=\mathcal{N}^{+}_{\text{bdy}}(\hat\D)\frac{(-4)^n\Gamma \left(h'+n\right)}{n!\Gamma \left(2 h' +2n\right)}\int _{-i \infty }^{i \infty }\frac{dc}{2\pi i}\frac{\hat f(c,-n ) \hat f(-c,-n)}{(\hat\Delta -h')^2-c^2}\,. 
\ee
where we remind the reader,
\be
\hat f(c,\tau)=\frac{\Gamma\left(\frac{c+\Delta_1-h'}{2}\right) \Gamma\left(\frac{c+\Delta_2-h'}{2}\right) \Gamma \left(c+h'-\tau \right)}{2 \Gamma (c)}\,.
\ee
We get $\hat t^{+}_n$, by subtracting the contributions to $\tilde {\hat t}^{+}_n$ from lower  blocks $G_{\D_1+\D_2+2k}$ with $k<n$. The general extression is too tedious to present, so we just indicate how to obtain them recursively,
\be\label{recursion}
 \hat t_n^+ = \tilde {\hat t}_n^+- \sum_{k=0}^{n-1} \bigg[\frac{(-1)^{n-k}{(\D_1+k)_{n-k}} {(\D_2+k)_{n-k}}}{(n-k)!(\D_1+\D_2+1-h+2k)_{n-k}}\bigg]  \hat t_k^N\,,
\ee
and for $n=0$ it is simply $\hat t_0^+ = \tilde {\hat t}_0^+$\,.

For the Dirichlet case, the analogous relations are simply obtained from the Neumann case as shown below,
\begin{align}
\mathcal{N}^{-}_{\text{bdy}}(\hat\D)&=\Big[\mathcal{N}^{+}_{\text{bdy}}(\hat\D)\Big]_{\substack{\D_1 \to \D_1+1 \\ \D_2 \to \D_2 +1 }}\,,\\
\hat a_n^-  (\D_1,\D_2,\hat\D)&= \hat a_n^+  (\D_1+1,\D_2+1,\hat\D)\,,\\
\tilde {\hat t}_n^-  (\D_1,\D_2,\hat\D)&= \tilde {\hat t}_n^+  (\D_1+1,\D_2+1,\hat\D)\,.
\end{align}

\subsection{Bulk exchange diagram}\label{bulkWittdetails}

Here we give the details of the block expansion coefficients from the bulk exchange operators. The Witten diagram is given by \eqref{Wbulk} . The following are the poles in $\tau$:
\begin{enumerate}
\item {$\tau=\frac{h\pm c}{2}+m_1$, for $m_1 \in \mathbb{Z}$ and $m_1\ge 0$}

\item{$\tau=\pm\frac{\D_{12}}{2}-m_2$, for $m_2 \in \mathbb{Z}$ and $m_2 \ge 0$}
\end{enumerate}
and the poles in $c$ are,
\begin{enumerate}
\item {$c=\pm(\D-h)$}

\item{$c=\pm(\D_1+\D_2-h+2m_3)$ for $m_3 \in \mathbb{Z}$ and $m_3\ge 0$}

\item{$c=\pm(1-h+2m_4)$ for   $m_4 \in \mathbb{Z}$ and $m_4\ge 0$}

\item{$c=\pm(h-2\tau+2m_5)$ for   $m_5 \in \mathbb{Z}$ and $m_5\ge 0$}\,.
\end{enumerate}

In the bulk channel, we have  $\xi\to 0$\,. So the $\tau$  contour is closed on the right. First we will compute the normalization $\mathcal{N}^{\nu}_{\text{bulk}}(\D)$. To get this, first we take the residue at the poles $\tau=\frac{h\pm c}{2}+m$,
{
\begin{align}\label{taures}
	\underset{\tau=-\frac{h\pm c}{2}+m}{\text{Res}} & \big[W_{\text{bulk}}^\nu\big]=\mathcal{N}^{\nu}_{\text{bulk}}\int \frac{dc}{(2\pi i)^2} {\scalebox{1}{ $\frac{\Gamma (\frac{1\pm c-h}{2} ) \Gamma (\frac{\Delta _1+\Delta _2\pm c-h}{2} ) \Gamma (\frac{2m\pm c+h}{2} ) \Gamma (\frac{h\pm c+2 m-\text{$\Delta_{12} $}}{2} ) \Gamma \left(\frac{h\pm c+2 m-\text{$\Delta_{21} $}}{2} \right)}{(-1)^m 2 \times  m! \Gamma (\pm c) \left(c^2-(h-\Delta )^2\right) \Gamma \left(\frac{1\mp c-h-2 m}{2} \right) \Gamma (h\pm c+2 m)}$}}\nonumber\\
& {\scalebox{0.85}{ $\times    f\big(-c,\frac{c+h+2 m}{2} \big) \left(4\xi\right)^{\frac{h\pm c-\text{$\Delta_1 $}-\text{$\Delta_2 $}+2 m}{2} }$}}\bigg[1+\nu\frac{{}_2F_1\big[{\scalebox{0.9}{ $\frac{1- c-h}{2} ,\frac{1+c-h}{2} ,\frac{1\mp c-h-2 m}{2} ,z$}}\big]}{\big(1-z\big)^{h-\frac{\text{$\Delta_1 $}+\text{$\Delta_2 $}+1}{2}}}\bigg]\,.
	\end{align}
}%
Let us remind the reader that,
\be
f(c,\tau)=\frac{\G(\frac{\D_1+\D_2-h+c}{2})\G(\frac{1+c-h}{2})\G(\frac{h+c}{2}-\tau)}{2\G(c)}\,.
\ee
The normalization $\mathcal{N}^{\nu}_{\text{bulk}}(\D)$ comes from requiring  coefficient of $G_{\D}$ in the bulk channel decomposition to be 1. In the $\xi\to 0$ expansion of $G_{\D}((1+\xi)^{-1}|\D_1,\D_2)$ the first term is $\xi^{\frac{\D-\D_1-\D_2}{2}}$. This is obtained from  \eqref{taures}, by putting $m=0$ and taking the residue at $c=\D-h$ ($c=h-\D$) for the upper (lower) sign with the $c$-contour closed on right (left). This gives,
\be\label{physN1}
\mathcal{N}^{\nu}_{\text{bulk}}(\D)={\scalebox{1.15}{ $\frac {2^{\D_1+\D_2-\D+2} \times \Gamma (\Delta )  \Gamma (\Delta -h+1)}{\Gamma \left(\frac{\Delta }{2}\right) \Gamma \left(\frac{ \Delta +\Delta _1-\Delta _2}{2}\right) \Gamma \left( \frac{\Delta -\Delta _1+\Delta _2}{2}\right) \Gamma \left(\frac{\Delta _1+\Delta _2 -\Delta }{2}\right) \Gamma \left(\frac{\Delta +1-d}{2}\right) \Gamma \left(\frac{\Delta +\Delta _1+\Delta _2-d}{2}\right)}$}}\,.
\ee
We will now look at the coeffients of the blocks $G_{\D_1+\D_2+2n}$. For simplicity of the expressions we will take $\D_1=\D_2=\D_\phi$, which is the case we are interested in anyway. First we have to obtain the power of $\xi^n$. This is done again from \eqref{taures} with the following steps\\

\textit{Step} 1. For the first term in parantheses, for the upper (lower) sign, we take the residue at $c=\D_1+\D_2-h+2(n-m)$  ($c=-\D_1-\D_2+h-2(n-m)$) and do the sums over $m=0,1,\cdots, n$\,. Finallly we expand in $\xi$ and the get the coefficient of $\xi^n$\,. 

\textit{Step} 2. For the second term (i.e. the hypergeometric) we expand in $\xi$, and integrate the coefficient of $\xi^n$ over $c$ using residues. Then we sum over all $m\ge 0 $.\\ \\ 
Let us denote the total coefficient of $\xi^n$ obtained in this way by  $\tilde t_n^{\nu}$. It can be expressed as an infinite sum of $_3F_2$ hypergeometric functions, but their expression is too tedious to display here.
The coefficient of the block $t^{\nu}_n$ is then obtained by using a recursion formula like \eqref{recursion}:
\be
- t^{\nu}_n =  \tilde t^{\nu}_n +  \sum_{k=0}^{n-1} \bigg[\frac{(-1)^{n-k}({(\D_\phi+k)_{n-k}})^2}{(n-k)!(2\D_\phi+1-h+2k)_{n-k}}\bigg]  t_k^\nu \,.
\ee

Let us now calculate the boundary channel decomposition coefficient $a_n(\D_1,\D_2,\D)$ of the boundary block $G_{\D_1+2n+\frac{\nu+1}{2}}$ from \eqref{Wbbulkdeco1}\,. The leading term of the block is $\xi^{-\D_1-2n-\frac{1-\nu}{2}}$ (times the prefactor $(1+\xi)^{\frac{\D_1+\D_2}{2}}$ coming from \eqref{oldnewblock})\,. Since the limit is $\xi\to \infty$, the $\tau$ contour is closed on the left. A certain power $\xi^{-\D_1-m}$ is obtained by taking residues of $\tau=\frac{\D_{21}}{2}-k$, with $k=0,1,\cdots m$, then expanding the terms at large  $\xi$ for each $k$ and summing. Let us call this coefficient $\tilde{a}_{m}^\nu $, where 
\begin{align}
\tilde{a}_{m}^\nu=\sum_{k=0}^m\Bigg[ & \frac{\mathcal{N}_{\text{bulk}}^{\nu}}{2^{2\D_1+2k}}  \int_{-i\infty}^{+i\infty} \frac{dc}{2\pi i}  \bigg[  \bigg(\d_{km}+\nu {\scalebox{0.9}{ $ \sum _{p=0}^{m-k} \binom{h-\frac{\D_1+\D_2+1}{2}}{m-k-p} \Big(\sum _{q=1}^p \frac{\binom{-q}{p-q} \left(\frac{1-c-h}{2} \right)_q \left(\frac{1+c-h}{2} \right)_q}{q! \left(k+\frac{\D_{12}+1}{2} \right)_q}+\delta _{p,0}\Big)  $}} \bigg) \nonumber\\
&\frac{f(c,\frac{\D_{21}}{2}-k)f(-c,\frac{\D_{21}}{2}-k)}{((\D-h)^2-c^2)} \bigg] \times  \frac{(-1)^k\G(\frac{\D_{21}}{2}-k)\G(\D_{21}-k)}{k!\G(\frac{1-\D_{21}}{2}+k)\G(\D_{21}-2k)} \Bigg]\,.
\end{align}
Integrals of this type can be evaluated using the following identity \cite{Gopakumar:2018xqi}
\begin{align}
\int_{-i \infty}^{i \infty}\frac{dc}{2\pi i}\frac{\prod_{i=1}^3\G(a_i-\frac{c}{2})\G(a_i+\frac{c}{2})}{(4a_4^2-c^2)\G(c)\G(-c)}&=\frac{\Gamma \left(a_1+a_2\right) \Gamma \left(a_1+a_3\right) \Gamma \left(a_2+a_4\right) \Gamma \left(a_3+a_4\right)}{\left(a_1+a_4\right) \Gamma \left(2 a_4+1\right)}\nonumber \\
\times & \, _3F_2\bigg( {\scalebox{1}{ $ \!\!\!\!\!\!\begin{matrix} & a_1+a_4,  a_4-a_2+1 , a_4-a_3+1 \\ &
 a_1+a_4+1, \ \   2 a_4+1\end{matrix};1  $}} \bigg)\,.
\end{align}
This gives the following,
\begin{align}
   \tilde a_m^\nu (\D_1, & \D_2,\D)  = \sum_{k=0}^m \frac{\mathcal{N}_{\text{bulk}}^{\nu} \Gamma \left(\frac{\Delta _{21}}{2}-k\right) \Gamma \left(\frac{\Delta +\Delta _1+\Delta _2}{2} - h\right)  \Gamma \left(k+\frac{\Delta -\Delta _{21}}{2} \right) }{2^{2\D_1+2k+1} (-1)^k k! \left(\Delta _{21}- k\right)_{-k} \Gamma (\Delta +1-h)  } \Bigg[ {\scalebox{0.9}{ $ \bigg( \delta _{k,m}+\nu  \binom{h-\frac{\D_1+\D_2+1}{2} }{m-k} \bigg)  $}}\nonumber \\& \times \frac{  \Gamma \left(\frac{\Delta _1+\Delta _2+1-d}{2} \right) }{(\Delta -d+1) } \, _3F_2 {\scalebox{0.75}{ $\bigg(  \!\!\!\!\!\!\begin{matrix} &\frac{\Delta +1-d}{2}, \frac{\Delta-\D_1-\D_2 }{2}+1 , \frac{\Delta+\D_{21} -d}{2}+1-k  \\ &
 \frac{\Delta+3 -d}{2} , \ \  \Delta -h+1 \end{matrix};1   \bigg)$}} + \nu \sum _{p=1}^{m-k} \sum _{q=1}^p {\scalebox{1}{ $  \binom{h-\frac{\D_1+\D_2+1}{2}}{m-k-p}  $}}  \nonumber \\ 
\times & {\scalebox{1}{ $  \bigg[ \frac{\binom{-q}{p-q}   \left(k-\frac{\Delta _{21}}{2}+\frac{1}{2}\right)_q \Gamma \left(\frac{\Delta _1+\Delta _2+1}{2}-h+q \right)  }{q! \left(k+\frac{\D_{12}+1}{2} \right)_q  (\Delta -2 h+2 q+1)}  $}} \, _3F_2 {\scalebox{0.75}{ $ \bigg(  \!\!\!\!\!\!\begin{matrix} &\frac{\Delta+1-d }{2}+q, \frac{\Delta-\D_1-\D_2 }{2}+1 , \frac{\Delta+\D_{21}-d }{2}+1-k  \\ &
 \frac{\Delta +3 -d}{2} +q, \ \  \Delta -h+1 \end{matrix};1 \bigg)     $}}    {\scalebox{0.8}{ $ \Bigg] $}} \Bigg].
\end{align}
The $a_n^\nu$-s are obtained by subtracting from $a_{2n+\frac{1-\nu}{2}}^\nu$ the  coefficients of the lower boundary blocks $a_{m<n}^{\nu}$. We give $a_{n}^{\nu}$ in the recursion formula,
\be\label{recursionbulk}
\small
a_n^\nu (\D_1,\D_2,\D)= \tilde{a}_{2n+\frac{1-\nu}{2}}^\nu - \sum_{k=0}^{n-1} \bigg[\frac{{(\D_1+2k')_{2n-2k}} {(\D_1-h+2k'+1)_{2n-2k}}}{(2n-2k)!(2\D_1+4k'+2-d)_{2n-2k}}\bigg] a_k^\nu(\D,\D_1,\D_2)\,,
\ee
where $k'=k+\frac{1-\nu}{4}$\,.

\subsection{Contact diagram}\label{contdetails}	

The contact Witten diagram \eqref{ContW} is rather easy to decompose into boundary and bulk blocks. There is only the $\tau$-integral, whose contour is closed on the right for boundary channel, and on the left for bulk channel. As we expand around $\xi\to\infty$ and $\xi\to 0$ the coefficients of $\xi^{-\D_1-n}$ and $\xi^n$ are respectively given by,
\begin{align}
\tilde d^\nu_n(\D_1,\D_2)&=\frac{(-1)^{n}  \Gamma (\Delta _{21}-n) \Gamma (n+\frac{1-\nu }{2}+\Delta _1)}{n! \Gamma \left(\frac{\Delta _{21}+1-2n}{2} \right)}\,, \nonumber \\  \text{and \ }  \tilde f^\nu_n(\D_1,\D_2)&=\frac{\Gamma \left(n+\frac{1-\nu }{2}+\Delta _1\right) \Gamma \left(n+\frac{1-\nu }{2}+\Delta _2\right)}{n! \Gamma \left(\frac{2 n+\Delta _1+\Delta _2+1}{2}\right)} \,.
\end{align}
From the above, obtaining the block decomposition coefficients is straightforward. For the Neumann case, we have the boundary channel expansion coefficients,
\be
{\scalebox{1}{ $d^+_n(\D_1,\D_2)=\frac{ \left(\Delta _1+\Delta _2+1-d\right) \Gamma \left( 2n+\Delta _1\right) \Gamma \left(\Delta _{21}- 2n\right) \Gamma \left(n+\frac{1}{2}+\Delta _1-h\right)  \left(\frac{\Delta _1+\Delta _2+3-d}{2}\right)_{n-1}}{  2^{2n+1}(-1)^{n}n! \Gamma \left(\frac{1+\D_{21}}{2} - n\right) \Gamma \left( 2n+\Delta _1+\frac{1}{2}-h\right)}$}}\,,
\ee
The corresponding coefficients for Dirichlet are simply given by 
\be
d_n^-(\D_1,\D_2)=d_n^+(\D_1+1,\D_2+1)\,.
\ee
The bulk decomposition coefficients, for the Neumann case, are given by,
\be
f_n^+(\D_1,\D_2)=\frac{(-1)^{n} (d-\text{$\Delta_1 $}-\text{$\Delta_2 $}-1) \Gamma (n+\text{$\Delta_1 $}) \Gamma (n+\text{$\Delta_2 $}) \left( \frac{\D_1+\D_2+3}{2}-h\right)_{n-1}}{2 (n!) \Gamma \left( n+\frac{\D_1+\D_2}{2}+\frac{1}{2}\right) \left(n+\text{$\Delta_1 $}+\text{$\Delta_2 $}-h\right)_n}\,.
\ee
For Dirichlet the corresponding coefficients $f_n^-$ can be obtained from $\tilde  f_n^-$, which are given by,
\be
\tilde f_{n}^-=\tilde f_{n}^+(\D_1+1,\D_2+1)\,.
\ee
One can approach the equal scalar limit, $\D_1\to \D_\phi+\frac{\D_{12}}{2}$ and $\D_2\to\D_\phi-\frac{\D_{12}}{2}$, to get
\bea
d_n^{\nu}(\Delta_1,\Delta_2)\overset{\Delta_{12}\to 0}=\frac{ e_n^{\nu}}{\Delta_{12}}+\frac 12 d_n^{\nu}+O(\Delta_{12}).
\eea
For later use, let us show these quantities for the Neumann case. Choosing the normalization $e_n=e_n^+/e_0^+$, \ $d_n=d_n^+/e_0^+$ and $f_n=f_n^+/e_0^+$, we get the following,
\be
e_n=\frac{16^{-n} (\Delta_\phi )_{2n} \Gamma \left(n+\Delta_\phi -h+\frac{1}{2}\right)^2}{ \Gamma (n+1)^2 \Gamma \left(\Delta_\phi -h+\frac{1}{2}\right) \Gamma \left(2 n+\Delta_\phi -h+\frac{1}{2}\right)}\,, \ \  \text{and } \  d_n=\frac{1}{2}\partial_n e_n\,.
\ee
The  bulk expansion coefficients are given by,
\be
f_n=\frac{(-1)^{n}\sqrt{\pi} (2\D_\phi+1-d) \Gamma (n+\text{$\Delta_\phi $})^2 \left( \D_\phi-h+\frac{3}{2}\right)_{n-1}}{2 (n!) \Gamma(\D_\phi)\Gamma \left( n+\D_\phi+\frac{1}{2}\right) \left(n+2\text{$\Delta_\phi $}-h\right)_n}\,.
\ee

\small
\parskip=-10pt
\bibliography{Functionals}
\bibliographystyle{jhep}

\end{document}